%Paper: hep-th/9211040
%From: warner@faraday.usc.edu (Nicholas Philip Warner)
%Date: Mon, 9 Nov 92 12:40:24 GMT-0800
%Date (revised): Tue, 8 Dec 92 20:25:04 GMT-0800

%
%%%%%%%%%%%%%%%%%%%%%%%%%%%%%%%%%%%%%%%%%%%%%%%%%%%%%%%%%%%%%
\input harvmac % needs version 9/91
\input tables
%%%%%%%%%%%%%%%%%%%%%%%%%%%%%%%%%%%%%%%%%%%%%%%%%%%%%%%%%%%%%
%
\newif\ifdraft

\noblackbox
\catcode`\@=11
\newif\iffrontpage
%
%%%%%%%%%%%%%%%%%%%%%%%%%%%%%%%%%%%%%%%%%%%%%%%%%%%%%%%%%%%%%%
%%%%% sizes, offsets etc
%%%%%%%%%%%%%%%%%%%%%%%%%%%%%%%%%%%%%%%%%%%%%%%%%%%%%%%%%%%%%%
\ifx\answ\bigans
\def\titleft{\titsm}
\magnification=1200\baselineskip=15pt plus 2pt minus 1pt
%
%%%%% unreduced mode: %%%%
%\voffset=0.35truein\hoffset=0.250truein
\advance\hoffset by-0.075truein
\hsize=6.15truein\vsize=600.truept\hsbody=\hsize\hstitle=\hsize
\else\let\lr=L
\def\titleft{\titla}
\magnification=1000\baselineskip=14pt plus 2pt minus 1pt
%
%%%%% reduced mode: %%%%%%%
%\hoffset=-.5truein\voffset=-.1truein
\vsize=6.5truein
\hstitle=8truein\hsbody=4.75truein
\fullhsize=10truein\hsize=\hsbody
\fi
\parskip=4pt plus 15pt minus 1pt
%%%%%%%%%%%%%%%%%%%%%%%%%%%%%%%%%%%%%%%%%%%%%%%%%%%%%%%%%%%%%%
%%%%%  fonts
%%%%%%%%%%%%%%%%%%%%%%%%%%%%%%%%%%%%%%%%%%%%%%%%%%%%%%%%%%%%%%
%%%%%%%%%%%%%%%%%%%%%%%%%%%%%%%%%%%%%%%%%%%%%%%%%%%%%%%%%%%%%%

\font\titla=cmr10 scaled\magstep3
\font\tenmss=cmss10
\font\absmss=cmss10 scaled\magstep1
\newfam\mssfam
\font\footrm=cmr8  \font\footrms=cmr5
\font\footrmss=cmr5   \font\footi=cmmi8
\font\footis=cmmi5   \font\footiss=cmmi5
\font\footsy=cmsy8   \font\footsys=cmsy5
\font\footsyss=cmsy5   \font\footbf=cmbx8
\font\footmss=cmss8
\def\footfont{\def\rm{\fam0\footrm}
\textfont0=\footrm \scriptfont0=\footrms
\scriptscriptfont0=\footrmss
\textfont1=\footi \scriptfont1=\footis
\scriptscriptfont1=\footiss
\textfont2=\footsy \scriptfont2=\footsys
\scriptscriptfont2=\footsyss
\textfont\itfam=\footi \def\it{\fam\itfam\footi}
\textfont\mssfam=\footmss \def\mss{\fam\mssfam\footmss}
\textfont\bffam=\footbf \def\bf{\fam\bffam\footbf} \rm}
\def\tenpoint{\def\rm{\fam0\tenrm}
\textfont0=\tenrm \scriptfont0=\sevenrm
\scriptscriptfont0=\fiverm
\textfont1=\teni  \scriptfont1=\seveni
\scriptscriptfont1=\fivei
\textfont2=\tensy \scriptfont2=\sevensy
\scriptscriptfont2=\fivesy
\textfont\itfam=\tenit \def\it{\fam\itfam\tenit}
\textfont\mssfam=\tenmss \def\mss{\fam\mssfam\tenmss}
\textfont\bffam=\tenbf \def\bf{\fam\bffam\tenbf} \rm}
\ifx\answ\bigans\def\abstractfont{\tenpoint}\else
\def\abstractfont{\def\rm{\fam0\absrm}
\textfont0=\absrm \scriptfont0=\absrms
\scriptscriptfont0=\absrmss
\textfont1=\absi \scriptfont1=\absis
\scriptscriptfont1=\absiss
\textfont2=\abssy \scriptfont2=\abssys
\scriptscriptfont2=\abssyss
\textfont\itfam=\bigit \def\it{\fam\itfam\bigit}
\textfont\mssfam=\absmss \def\mss{\fam\mssfam\absmss}
\textfont\bffam=\absbf \def\bf{\fam\bffam\absbf}\rm}\fi
%
%%%%%%%%%%%%%%%%%%%%%%%%%%%%%%%%%%%%%%%%%%%%%%%%%%%%%%%%%%%%%%
%%%%% footnotes   (adapted from PHYZZX)
%%%%%%%%%%%%%%%%%%%%%%%%%%%%%%%%%%%%%%%%%%%%%%%%%%%%%%%%%%%%%%
\def\f@@t{\baselineskip10pt\lineskip0pt\lineskiplimit0pt
\bgroup\aftergroup\@foot\let\next}
\setbox\strutbox=\hbox{\vrule height 8.pt depth 3.5pt width\z@}
\def\vfootnote#1{\insert\footins\bgroup
\baselineskip10pt\footfont
\interlinepenalty=\interfootnotelinepenalty
\floatingpenalty=20000
\splittopskip=\ht\strutbox \boxmaxdepth=\dp\strutbox
\leftskip=24pt \rightskip=\z@skip
\parindent=12pt \parfillskip=0pt plus 1fil
\spaceskip=\z@skip \xspaceskip=\z@skip
\Textindent{$#1$}\footstrut\futurelet\next\fo@t}
\def\Textindent#1{\noindent\llap{#1\enspace}\ignorespaces}
\def\footnote#1{\attach{#1}\vfootnote{#1}}%

\def\foot{\attach\footsymbolgen\vfootnote{\footsymbol}}
\let\footsymbol=\star
\newcount\lastf@@t           \lastf@@t=-1
\newcount\footsymbolcount    \footsymbolcount=0
\def\footsymbolgen{\relax\footsym
\global\lastf@@t=\pageno\footsymbol}
\def\footsym{\ifnum\footsymbolcount<0
\global\footsymbolcount=0\fi
{\iffrontpage \else \advance\lastf@@t by 1 \fi
\ifnum\lastf@@t<\pageno \global\footsymbolcount=0
\else \global\advance\footsymbolcount by 1 \fi }
\ifcase\footsymbolcount \fd@f\star\or
\fd@f\dagger\or \fd@f\ast\or
\fd@f\ddagger\or \fd@f\natural\or
\fd@f\diamond\or \fd@f\bullet\or
\fd@f\nabla\else \fd@f\dagger
\global\footsymbolcount=0 \fi }
\def\fd@f#1{\xdef\footsymbol{#1}}
\def\space@ver#1{\let\@sf=\empty \ifmmode #1\else \ifhmode
\edef\@sf{\spacefactor=\the\spacefactor}
\unskip${}#1$\relax\fi\fi}
\def\attach#1{\space@ver{\strut^{\mkern 2mu #1}}\@sf}
%
%%%%%%%%%%%%%%%%%%%%%%%%%%%%%%%%%%%%%%%%%%%%%%%%%%%%%%%%%%%%%%
%%%%% References
%%%%%%%%%%%%%%%%%%%%%%%%%%%%%%%%%%%%%%%%%%%%%%%%%%%%%%%%%%%%%%
\newif\ifnref
\def\rrr#1#2{\relax\ifnref\nref#1{#2}\else\ref#1{#2}\fi}
\def\ldf#1#2{\begingroup\obeylines
\gdef#1{\rrr{#1}{#2}}\endgroup\unskip}
\def\nrf#1{\nreftrue{#1}\nreffalse}
\def\doubref#1#2{\refs{{#1},{#2}}}
\def\multref#1#2#3{\nrf{#1#2#3}\refs{#1{--}#3}}
\nreffalse
\def\refout{\listrefs}
%
%%%%%%%%%%%%%%%%%%%%%%%%%%%%%%%%%%%%%%%%%%%%%%%%%%%%%%%%%%%%%%
%%%%%%% eq numbering
%%%%%%%%%%%%%%%%%%%%%%%%%%%%%%%%%%%%%%%%%%%%%%%%%%%%%%%%%%%%%%
\def\EQN#1{\xdef #1{(\secsym\the\meqno)}
\writedef{#1\leftbracket#1}%
\global\advance\meqno by1\eqno#1\eqlabeL#1}
\def\eqnalign#1{\xdef #1{(\secsym\the\meqno)}
\writedef{#1\leftbracket#1}%
\global\advance\meqno by1#1\eqlabeL{#1}}
%
%%%%%%%%%%%%%%%%%%%%%%%%%%%%%%%%%%%%%%%%%%%%%%%%%%%%%%%%%%%%%%
%%%%%%  macros for titlepage, marginnotes, etc
%%%%%%%%%%%%%%%%%%%%%%%%%%%%%%%%%%%%%%%%%%%%%%%%%%%%%%%%%%%%%%
\def\chap#1{\newsec{#1}}
\def\chapter#1{\chap{#1}}
\def\sect#1{\subsec{{ #1}}}
\def\section#1{\sect{#1}}
\def\\{\ifnum\lastpenalty=-10000\relax
\else\hfil\penalty-10000\fi\ignorespaces}
\def\note#1{\leavevmode%
\edef\@@marginsf{\spacefactor=\the\spacefactor\relax}%
\ifdraft\strut\vadjust{%
\hbox to0pt{\hskip\hsize%
\ifx\answ\bigans\hskip.1in\else\hskip .1in\fi%
\vbox to0pt{\vskip-\dp
%\vskip4pt
\strutbox\sevenbf\baselineskip=8pt plus 1pt minus 1pt%
\ifx\answ\bigans\hsize=.7in\else\hsize=.35in\fi%
\tolerance=5000 \hbadness=5000%
\leftskip=0pt \rightskip=0pt \everypar={}%
\raggedright\parskip=0pt \parindent=0pt%
\vskip-\ht\strutbox\noindent\strut#1\par%
\vss}\hss}}\fi\@@marginsf\kern-.01cm}
\def\titlepage{%
\frontpagetrue\nopagenumbers\abstractfont%
\hsize=\hstitle\rightline{\vbox{\baselineskip=10pt%
{\abstractfont\pubnum}}}\pageno=0}
\frontpagefalse
\def\pubnum{}
\def\pdate{\number\month/\number\yearltd}
\def\makefootline{\iffrontpage\vskip .27truein
\line{\the\footline}
%\vskip -.1truein\line{\pdate\hfil}
\vskip -.1truein\leftline{\vbox{\baselineskip=10pt%
{\abstractfont\pdate}}}
\else\vskip.5cm\line{\hss \tenrm $-$ \folio\ $-$ \hss}\fi}
\def\title#1{\vskip .7truecm\titlestyle{\titleft #1}}
\def\titlestyle#1{\par\begingroup \interlinepenalty=9999
\leftskip=0.02\hsize plus 0.23\hsize minus 0.02\hsize
\rightskip=\leftskip \parfillskip=0pt
\hyphenpenalty=9000 \exhyphenpenalty=9000
\tolerance=9999 \pretolerance=9000
\spaceskip=0.333em \xspaceskip=0.5em
\noindent #1\par\endgroup }
\def\autskip{\ifx\answ\bigans\vskip.5truecm\else\vskip.1cm\fi}
\def\author#1{\vskip .7in \centerline{#1}}

\def\address#1{\ifx\answ\bigans\vskip.2truecm
\else\vskip.1cm\fi{\it \centerline{#1}}}
\def\abstract#1{
\vskip .5in\vfil\centerline
{\bf Abstract}\penalty1000
{{\smallskip\ifx\answ\bigans\leftskip 2pc \rightskip 2pc
\else\leftskip 5pc \rightskip 5pc\fi
\noindent\abstractfont \baselineskip=12pt
{#1} \smallskip}}
\penalty-1000}
\def\endpage{\tenpoint\supereject\global\hsize=\hsbody%
\frontpagefalse\footline={\hss\tenrm\folio\hss}}
%

%%%%%%%%%%%%%%%%%%%%%%%%%%%%%%%%%%%%%%%%%%%%%%%%%%%%%%%%%%%%%%

%%%%%%%%%%%%%%%%%%%%%%%%%%%%%%%%%%%%%%%%%%%%%%%%%%%%%%%%%%%%%%
\def\bfone{\relax{\rm 1\kern-.35em 1}}
\def\inbar{\vrule height1.5ex width.4pt depth0pt}
\def\IC{\relax\,\hbox{$\inbar\kern-.3em{\mss C}$}}
\def\ID{\relax{\rm I\kern-.18em D}}
\def\IF{\relax{\rm I\kern-.18em F}}
\def\IH{\relax{\rm I\kern-.18em H}}
\def\II{\relax{\rm I\kern-.17em I}}
\def\IN{\relax{\rm I\kern-.18em N}}
\def\IP{\relax{\rm I\kern-.18em P}}
\def\IQ{\relax\,\hbox{$\inbar\kern-.3em{\rm Q}$}}
\def\IR{\relax{\rm I\kern-.18em R}}
\font\cmss=cmss10 \font\cmsss=cmss10 at 7pt
\def\ZZ{\relax\ifmmode\mathchoice
{\hbox{\cmss Z\kern-.4em Z}}{\hbox{\cmss Z\kern-.4em Z}}
{\lower.9pt\hbox{\cmsss Z\kern-.4em Z}}
{\lower1.2pt\hbox{\cmsss Z\kern-.4em Z}}\else{\cmss Z\kern-.4em Z}\fi}
\def\a{\alpha} \def\b{\beta} \def\d{\delta}
 
 \def\l{\lambda}

\def\cF{{\cal F}} 
 
\def\cJ{{\cal J}} 
 \def\cM{{\cal M}}
 \def\cO{{\cal O}}
\def\cP{{\cal P}} \def\cQ{{\cal Q}}
\def\cR{{\cal R}} \def\cV{{\cal V}}
\def\nup#1({Nucl.\ Phys.\ $\us {B#1}$\ (}
\def\plt#1({Phys.\ Lett.\ $\us  {#1}$\ (}
\def\cmp#1({Comm.\ Math.\ Phys.\ $\us  {#1}$\ (}
\def\prp#1({Phys.\ Rep.\ $\us  {#1}$\ (}
\def\prl#1({Phys.\ Rev.\ Lett.\ $\us  {#1}$\ (}
\def\prv#1({Phys.\ Rev.\ $\us  {#1}$\ (}
\def\mpl#1({Mod.\ Phys.\ Let.\ $\us  {A#1}$\ (}
\def\ijmp#1({Int.\ J.\ Mod.\ Phys.\ $\us{A#1}$\ (}
\def\tit#1|{{\it #1},\ }
%
%%%%%%%%%%%%%%%%%%%%%%%%%%%%%%%%%%%%%%%%%%%%%%%%%%%%%%%%%%%%%%
%%%%% misc macros %%%%%
%%%%%%%%%%%%%%%%%%%%%%%%%%%%%%%%%%%%%%%%%%%%%%%%%%%%%%%%%%%%%%

%
\def\refout{\vskip2.truecm\immediate\closeout\rfile\writestoppt
\baselineskip=14pt\centerline{{\bf References}}\bigskip{\frenchspacing%
\parindent=20pt\escapechar=` \input \jobname.refs\vfill\eject}\nonfrenchspacing}

\def\tilde{\widetilde}
\def\bar{\overline}
\def\us#1{\underline{#1}}

\def\hat{\widehat}

\def\Coe#1.#2.{{#1\over #2}}
\def\coeff#1#2{\relax{\textstyle {#1 \over #2}}\displaystyle}
\def\coe#1.#2.{\relax{\textstyle {#1 \over #2}}\displaystyle}
\def\half{{1 \over 2}}
\def\shalf{\relax{\textstyle {1 \over 2}}\displaystyle}

\def\to{\rightarrow}
\def\notin{\hbox{{$\in$}\kern-.51em\hbox{/}}}
\def\shdot{\!\cdot\!}
\def\ket#1{\,\big|\,#1\,\big>\,}

\def\exx#1{e^{{\displaystyle #1}}}
\def\del{\partial}

\def\nex#1{$N\!=\!#1$}

\catcode`\@=12
%%%%%%%%%%%%%%%%%%%%%%%%%%%%%%%%%%%%%%%%%%%%%%%%%%%%%%%%%%%%%%%%%%%%%
%%%%%%%%%%%%%%%%%%%%%%%%%%%%%%%%%%%%%%%%%%%%%%%%%%%%%%%%%%%%%%%%%%%%%
%%%%%%%%%%%%%%%%%%%%%%%%%%%%%%%%%%%%%%%%%%%%%%%%%%%%%%%%%%%%%%%%%%%%%
%
\ldf\TOPALG{E.\ Witten, \cmp{117} (1988) 353; \cmp{118} (1988) 411;
\nup340 (1990) 281; T.\ Eguchi and S.\ Yang, \mpl4 (1990) 1693.}
\ldf\ssh{A.\ Schwimmer and N.\ Seiberg, \plt 184B (1987) 91.}
\ldf\VVtopgr{E.\ Verlinde and H.\ Verlinde, \nup348 (1991) 457.}
\ldf\Muss{G.\ Mussardo, G.\ Sotkov, M.\ Stanishkov, Int.\ J.\ Mod.\ Phys.\
A4 (1989) 1135; N.\ Ohta and H.\ Suzuki, \nup332(1990) 146.}
\ldf\spieg{M.\ Spiegelglas and S.\ Yankielowicz, {\it G/G topological field
theories by cosetting G(k)}, preprint TECHNION-PH-90-34.}
\ldf\Witopgr{E.\ Witten,  \nup340 (1990) 281. }
\ldf\Witbh{E.\ Witten,   Phys.\ Rev.\ D44 (1991) 314. }
\ldf\Witgg{E.\ Witten,  \nup371(1992)191. }
\ldf\Witgr{E.\ Witten, \nup373 (1992) 187. }
\ldf\OV{H.\ Ooguri and C.\ Vafa, \mpl5 (1990) 1389; \nup361(1991) 469.}
\ldf\elias{I.\ Bakas and E.\ Kiritsis,  {\it Beyond the large N limit:
Nonlinear $W_\infty$ as symmetry of the
SL(2,R)/U(1) coset model}, preprint UCB-PTH-91-44.}
\ldf\BLNW{M.\ Bershadsky, W.\ Lerche, D.\ Nemeschansky and N.P.\ Warner,
\plt B292 (1992) 35.}
\ldf\FLMW{P.\ Fendley, W.\ Lerche, S.\ Mathur and N.P.\ Warner, \nup348 (1991)
66. }
\ldf\FMVW{P.\ Fendley, S.\ Mathur C.\ Vafa and N.P.\ Warner, \nup348 (1991) 66.
}
\ldf\LeWa{W.\ Lerche and N.P.\ Warner, \nup358 (1991) 571. }
\ldf\Loss{A. Lossev, {\it Descendants constructed from matter field and K.
Saito higher residue pairing in Landau-Ginzburg theories coupled to topological
gravity},
preprint TPI-MINN-92-40-T. }
\ldf\LNW{W.\ Lerche, D.\ Nemeschansky and N.P.\ Warner, unpublished.}
\ldf\DLP{L.\ Dixon, M.\ Peskin and J.\ Lykken, \nup325(1989) 329.}
\ldf\BGS{B.\ Gato-Rivera and A.M.\ Semikhatov, \plt B293 (1992) 72.}
\ldf\BSS{E.\ Bergshoeff, A.\ Sevrin and X.\ Shen, {\it A Derivation of the BRST
operator for noncritical W strings}, preprint  UG-8-92.}
\ldf\LVW{W.\ Lerche, C.\ Vafa and N.P.\ Warner, \nup324 (1989) 427.}
\ldf\WL{W.\ Lerche, \plt252B (1990) 349.}
\ldf\cring{D.\ Gepner, {\it A comment on the chiral algebras of quotient
superconformal field
theories}, preprint PUPT-1130; S.\ Hosono and A.\ Tsuchiya,
\cmp136(1991) 451.}
\ldf\PolyA{A.\ Polyakov, \plt B103(1981) 207. }
\ldf\LZ{B.\ Lian and G.\ Zuckerman, \plt254B (1991) 417, \plt266B(1991)21,
\cmp145 (1992) 561.}
\ldf\EXTRA{A.\ Polyakov, \mpl6(1991) 635;
I.\ Klebanov and A.\ Polyakov, \mpl6(1991) 3273;
S.\ Mukherji, S.\  Mukhi and A.\ Sen, \plt 266B (1991) 337;
H.\  Kanno and M.\  Sarmadi, preprint IC/92/150;
K.\ Itoh and N.\ Ohta, \nup377 (1992) 113;
N.\ Chair, V.\ Dobrev and H.\ Kanno, \plt283B (1992) 194;
S.\ Govindarajan, T.\ Jayamaran, V.\ John and P.\ Majumdar,
\mpl7 (1992) 1063; S.\ Govindarajan, T.\ Jayamaran and V.\ John, {\it Chiral
rings and physical states in $c<1$ string
theory}, preprint IMSc-92/30.}
\ldf\NY{D.\ Nemeschansky and S.\ Yankielowicz, {\it N=2 W-algebras,
Kazama-Suzuki models and Drinfeld-Sokolov reduction}, preprint USC-91-005A. }
\ldf\NW{D.\ Nemeschansky and N.P.\ Warner, \nup 380(1992) 241.}
\ldf\EY{T.\ Eguchi, S.\ Hosono and S.K.\ Yang, \cmp140(1991) 159;
T.\ Eguchi, T.\ Kawai, S.\ Mizoguchi and S.K.\ Yang, {\it Character formulas
for coset N=2 superconformal theories}, preprint KEK-TH-303.}
\ldf\isr{O.\ Aharony, O.\ Ganor, J.\ Sonnenschein and  S.\ Yankielowicz,
preprint TAUP-1961-92; O.\ Aharony, J.\ Sonnenschein and  S.\ Yankielowicz,
\plt B289 (1992) 309 }
\ldf\BMP{P.\ Bouwknegt, J.\ McCarthy and K.\ Pilch, \cmp145(1992) 541; {\it
Fock space resolutions of the Virasoro highest weight modules with $c\leq1$},
preprint CERN-TH.6196/91; {\it BRST analysis of physical states for 2d (super)
gravity coupled to (super) conformal matter}, preprint CERN-TH.6279/91.}
\ldf\BMPa{P.\ Bouwknegt, J.\ McCarthy and K.\ Pilch, Lett.\ Math.\ Phys.\
{\bf 23} (1991) 193.}
\ldf\BMPtop{P.\ Bouwknegt, J.\ McCarthy and K.\ Pilch, {\it On physical states
in
2d (topological) gravity}, preprint CERN-TH.6645/92.}
\ldf\Wit{E.\ Witten, \nup373(1992) 187.}
\ldf\Hull{C.\ Hull, \plt B240 (1990) 110; \nup353(1991) 707;
\plt259B(1991) 68.}
\ldf\ITO{K.\ Ito, \plt B259(1991) 73; \nup370(1992) 123.}
\ldf\larry{L.J.\ Romans, \nup369 (1992) 403.}
\ldf\bo{M.\ Bershadsky and H.\ Ooguri, \cmp126(1989) 49;  M.~Bershadsky and
H.~Ooguri, \plt{229B} (1989) 374.}
\ldf\topw{K.\ Li, \plt B251 (1990) 54, \nup346 (1990) 329; S.\ Hosono, {\it
Algebraic definition of topological W-gravity}, preprint UT-588-TOKYO.}
\ldf\TM{J.\ Thierry-Mieg, \plt197B(1987) 368.}
\ldf\FLZ{A.B.\ Zamolodchikov, Theor.\ Math.\ Phys.\ 65 (1985) 1205;
V.A.\ Fateev and A.B.\ Zamolodchikov, \nup280(1987) 644;
F.\ Bais, P.\ Bouwknegt, K.\ Schoutens and M.\ Surridge, \nup304(1988) 348;
V.A.\ Fateev and S.L.\ Luk'yanov, \ijmp3(1988) 507.}
\ldf\DK{J.\ Distler and T.\ Kawai, \nup321(1989) 509.}
\ldf\DiNe{J.\ Distler and P.\ Nelson, \cmp{138} (1991) 273.}
\ldf\LPSX{H.\ Lu, C.N.\ Pope, S.\ Schrans and K.\ Xu,
% {\it The complete Spectrum of the $W$-String},
preprint CTP-TAMU-5/92, KUL-TF-92/1, KUL-TF-92/1;
H.\ Lu, C.N.\ Pope, S.\ Schrans and X.J.\ Wang, preprint CTP-TAMU-15/92;
C.N.\ Pope, preprint CTP-TAMU-30/92.}
\ldf\MS{P.\ Mansfield and B.\ Spence, \nup362(1991) 294.}
\ldf\KS{Y.\ Kazama and H.\ Suzuki, \plt216B(1989)
112; \nup321(1989) 232.}
\ldf\Moro{A.\ Marshakov, A.\ Mironov, A.\ Morozov and M.\ Olshanetsky, preprint
FIAN/TD-02/92 and ITEP-M-2/92.}
\ldf\Keke{K.\ Li, \nup354(1991) 711.}
\ldf\Watts{G.\ Watts, \nup326(1989) 648.}
\ldf\kac{V.\ Dotsenko, \mpl6 (1991) 3601; Y.\ Kitazawa, \plt B265(1991) 262.}
\ldf\ind{
S.\ Govindarajan, T.\ Jayamaran, V.\ John and P.\ Majumdar, \mpl7 (1992) 1063;
S.\ Govindarajan, T.\ Jayamaran and V.\ John, {\it Chiral rings and physical
states in $c<1$ string theory}, preprint IMSc-92/30; {\it Genus zero
correlation functions in $c<1$ string theory}, preprint IMSC-92-35}
\ldf\Popring{C.N.\ Pope, E. Sezgin\ , K.S. Stelle\ and X.J.\ Wang, {\it
Discrete states in the $W_3$ string}, preprint CTP-TAMU-64-92.}
\ldf\modul{D.\ Kutasov, E.\ Martinec and N.\ Seiberg, \plt B276 (1992) 437.}
\ldf\dis{J.\ Distler, \nup342(1990) 523.}
\ldf\Ind{S.\ Govindarajan, T.\ Jayamaran, V.\ John and P.\
Majumdar, preprint IMSc-91/40.}
\ldf\Das{S. Das, A. Dhar and S. Kalyana Rama, preprint TIFR/TH/91-20.}
\ldf\Sei{N.\ Seiberg, Progr.\ Theor.\ Phys.\ Suppl.\ 102 (1990) 319.}
\ldf\texans{C.N.\ Pope, L.J.\ Romans and K.S.\ Stelle, \plt268B(1991) 167;
\plt269B(1991) 287.}
\ldf\FMS{D.\ Friedan, E.\ Martinec and S.\ Shenker, \nup271(1986) 93.}
\ldf\OPE{K. Thielemans, Int. J. Mod. Phys. C Vol. {\bf 2}, No. 3,
787 (1991).}
\ldf\BG{A.\ Bilal and J.\ Gervais, \nup326(1989) 222.}
\ldf\LPSX{H.\ Lu, C.N.\ Pope, S.\ Schrans and K.\ Xu,
% {\it The complete Spectrum of the $W$-String},
preprint CTP-TAMU-5/92, KUL-TF-92/1, KUL-TF-92/1;
H.\ Lu, C.N.\ Pope, S.\ Schrans and X.J.\ Wang, preprint CTP-TAMU-15/92;
C.N.\ Pope, preprint CTP-TAMU-30/92.}
\ldf\MS{P.\ Mansfield and B.\ Spence, \nup362(1991) 294.}
\ldf\Das{S. Das, A. Dhar and S. Kalyana Rama, preprint TIFR/TH/91-20.}
%

%%%%%%%%%%%%%%%%%%%%%%%%%%%%%%%%%%%%%%%%%%%%%%%%%%%%%%%%%%%%%%%%%%%%%
%%%%%%%%%%%%%%%%%%%%%%%%%%%%%%%%%%%%%%%%%%%%%%%%%%%%%%%%%%%%%%%%%%%%%
%%%%%%%%%%%%%%%%%%%%%%%%%%%%%%%%%%%%%%%%%%%%%%%%%%%%%%%%%%%%%%%%%%%%%

\def\brs{$BRST$\ }
\def\qbrs{\cQ_{BRST}}
\def\jbrs{\cJ_{BRST}}
\def\mm#1#2{\cM_{#1,#2}}
\def\mpq{\mm pq}

\def\bi#1{b_{#1}}
\def\ci#1{c_{#1}}
\def\zw#1#2.{{#2\over(z-w)^{#1}}}
\def\tg{T_{gh}}
\def\cW{{\cal W}}
\def\lv{{Liouville}}

\def\sc{superconformal\ }
\def\sca{\sc algebra\ }
\def\qt{\tilde\cQ}
\def\zmw#1{(z-w)^{#1}}

\def\prs{\Phi_{r_1,r_2;s_1,s_2}}
\def\rs{{r1,r2;s1,s2}}
\def\zl{^{(Z)}}
\def\xc{\,c_1(\del c_2)c_2\,}
\def\oz#1{\cO^{(#1)}}
\def\cW{{\cal W}}
\thicksize=0.pt
\def\g{{\gamma^0}}
\def\gn{\gamma^0}
\def\LG{Lan\-dau-Ginz\-burg\ }

%%%%%%%%%%%%%%%%%%%%%%%%%%%%%%%%%%%%%%%%%%%%%%%%%%%%%%%%%%%%%%%%%%%%%
%%%%%%%%%%%%%%%%%%%%%%%%%%%%%%%%%%%%%%%%%%%%%%%%%%%%%%%%%%%%%%%%%%%%%
%%%%%%%%%%%%%%%%%%%%%%%%%%%%%%%%%%%%%%%%%%%%%%%%%%%%%%%%%%%%%%%%%%%%%

%\draft
\baselineskip=15pt plus 2pt minus 1pt

\def\pubnum{
\hbox{CALT-68-1832}
\hbox{CERN-TH.6694/92}
\hbox{HUTP-A061/92}
\hbox{USC-92/021}
\hbox{hepth@xxx/9211040}
}
\def\pdate{
\hbox{CERN-TH.6694/92}
\hbox{October 1992}
}

\titlepage
\title
 {Extended N=2 Superconformal Structure of\break
 Gravity and W-Gravity Coupled to Matter}
\vskip-1.1cm\autskip
\author{M.\ Bershadsky}
\address{Lyman Laboratory, Harvard University, Cambridge, MA 02138}
\vskip-.7truecm
\author{W.\ Lerche}
\address{California Institute of Technology, Pasadena, CA 91125}
\address{and CERN, Geneva, Switzerland}
\vskip-.7truecm
\author{D.\ Nemeschansky and N.P.\ Warner}
\address{Physics Department, U.S.C., University Park,
Los Angeles, CA 90089}
\vskip-1.truecm
\abstract{We show that almost all string theories, including the bosonic
string, the superstring and $W$-string theories,  possess a twisted \nex2
superconformal symmetry.  This enables us to establish a connection between
topological gravity and the field theoretical description of  matter coupled to
gravity.  We also show how the \brs operators of the $W_n$-string
can be obtained by hamiltonian reduction  of $SL(n|n-1)$.  The tachyonic and
ground ring states of $W$-strings are described in the light of the \nex2
superconformal structure, and the ground ring generators for the non-critical
$W_3$-string are explicitly constructed.  The relationship to $G/G$ models and
quantum integrable systems is also briefly described.}

\endpage

%%%%%%%%%%%%%%%%%%%%%%%%%%%%%%%%%%%%%%%%%%%%%%%%%%%%%%%%%%%%%%%%%%%%%
%%%%%%%%%%%%%%%%%%%%%%%%%%%%%%%%%%%%%%%%%%%%%%%%%%%%%%%%%%%%%%%%%%%%%
%%%%%%%%%%%%%%%%%%%%%%%%%%%%%%%%%%%%%%%%%%%%%%%%%%%%%%%%%%%%%%%%%%%%%

\chap{Introduction}

During the past decade our understanding of string theory
and two-dimensional quantum gravity has grown considerably.
Much of the recent progress was stimulated by the success of
matrix models \ref\mat{D.J.~Gross and A.A.~Migdal, \prl{64} (1990) 717;
M.~Douglas and S.~Shenker, \nup{235} (1990) 635;
E.~Brezin and V.~Kazakov, \plt{236B} (1990) 144.}.  Still more recently,
Witten has shown that topological field theories can be used to describe
two-dimensional quantum gravity \Witopgr.  In this approach  the unperturbed
correlation  functions of the latter are identified with the intersection
numbers of cycles on the moduli space of Riemann surfaces. The matrix model
and the topological field theory approach are now known to be equivalent to
each other \ref\kon{M.~ Kontsevich, ``Intersection Theory on the Moduli Space
of Curves and the Matrix Airy Function,'' Max Plank Institute preprint
MPI/91-77}.

One  of the more important problems in two-dimensional string theory is
to reconstruct the results of topological gravity (or matrix models)
directly from the continuum, field theoretic approach in which matter fields
are directly coupled to the Liouville field.  In the topological formulation of
two-dimensional quantum gravity coupled to matter \Keke,  the relevant matter
fields come from a twisted \nex2 superconformal minimal model \TOPALG. In this
paper we establish a direct connection between the continuum formulation of
matter models coupled to gravity and such twisted \nex2 superconformal models.
Several authors have observed (see, for example, \ref\DVV{ R.~Dijkgraaf,
E.~Verlinde and H.~Verlinde, ``Notes on Topological String Theory and 2-D
Quantum Gravity,''  Lectures given at Spring School on Strings and Quantum
Gravity, Trieste, Italy, Apr 24 -- May 2, 1990 and at Cargese Workshop
on Random Surfaces, Quantum Gravity and Strings, Cargese, France, May
28 -- June 1, 1990.})  that the BRST current and the anti-ghost field, $b(z)$,
generate an algebra that  is reminiscent of, but apparently not identical to,
the \nex2 superconformal  algebra.  It does not, however, seem to be broadly
appreciated that the BRST current can be modified by total derivative terms so
that the anti-ghost and the physical \brs current exactly generate a
topologically twisted  \TOPALG\ \nex2 superconformal  algebra.
Indeed, there are several ways in which this modification can be done.  This
observation was first made in \BGS.   The underlying \nex2 superconformal
symmetry is quite generic, and is present in every string theory.

Standard two-dimensional gravity is moderately well understood, and thus
interest is now being focussed on other string theories for which the matrix
model discription, or a clear topological description, is either deficient or
completely lacking. This happens, for example, when the conformal matter system
has central charge larger than one. In these circumstances the conformal matter
system may have a larger symmetry, and it can often be coupled to some extended
background geometry. In addition to well known generalizations to
super-geometry, one can also try to define a ``$W$-geometry'', in which
conformal $W$-matter is coupled to $W$-gravity. We will show that all
$W$-gravity theories have a twisted \nex2 super-$W$ conformal symmetry. In
fact, this phenomenon is quite general: Gauging the local $W$-symmetry
automatically leads to the existence of an \nex2 supersymmetry. Perhaps more
importantly, one can reverse this perspective, and use a manifestly \nex2
superconformal hamiltonian reduction to obtain the the \brs structure for a
general $W$-gravity coupled to $W$-matter. In section 2 of this paper we will
explicitly construct the (extended) \nex2 superconformal generators for the
non-critical $W$-strings for which the \brs current is explicitly known. In
section 3 we will show how the general \brs current for non-critical
``$W_n$-strings'' can be obtained by hamiltonian reduction of the affine
super-Lie algebra $SL(n|n-1)$. This approach makes manifest, and directly
exploits, the \nex2 superconformal structure. In addition, hamiltonian
reduction establishes a direct relation between $W$-gravities and topological
field theories based upon Kazama-Suzuki models \KS\ obtained from
$SU(n)_k/U(n-1)$.

In section 4 of this paper we will describe the relationship between the
spectrum of the standard two-dimensional string and the \nex2 superconformal
minimal models.   Specifically, we show that the coupling  between topological
matter and topological gravity amounts to effectively undoing one of the \nex2
superconformal screening operations, thereby producing a field theory that is
equivalent to standard two-dimensional gravity.  We also discuss the
generalization of this to $W$-strings.  One can then use some of the machinery
of \nex2 superconformal field theories to obtain a new perspective on
$W$-string theory, and most particularly upon ground rings \Witgr. (The ground
ring and extra states for the non-critical $W_3$-string will be discussed in
Appendices A and B of this paper.)  There is, however, one subtlety: the
complete physical spectrum of the string theory generically forms a non-unitary
 representation of the \nex2 superconformal algebra.  As a result, some of the
standard theorems about chiral rings no longer apply.

When the \nex2 superconformal structure of string theories is combined with the
known relationship between \nex2 superconformal models and topological $G/G$
models \NW\ we also see how the continuum field theoretic approach to
two-dimensional quantum $W$-gravity is related to the approach based upon
topological $G/G$ models \multref\spieg\Witgg\isr. In particular, we argue that
the topological $G/G$ models and the standard continuum description of
$W$-gravity simply yield string theories at different values of the
cosmological constant. We also show that a non-critical $W$-string possesses an
infinite number of conserved $W$-charges, even in the presence of a
cosmological constant perturbation. These issues will be discussed in section
4.

There are also natural questions about further generalizations of this hidden
superconformal structure.  One obvious thing to consider is the \brs structure
of two-dimensional, \nex1 supergravity.  We discuss this in the next section
and show
that there is in fact a \nex3 superconformal symmetry.  We similarly expect
that the \nex2 supergravity theories \OV\ actually have an \nex4 superconformal
structure.  It would be very interesting to see whether this supersymmetric
extension of symmetry algebras has some analogue in higher dimensional field
theories.

%%%%%%%%%%%%%%%%%%%%%%%%%%%%%%%%%%%%%%%%%%%%%%%%%%%%%%%%%%%%%%%%%%%%%
%%%%%%%%%%%%%%%%%%%%%%%%%%%%%%%%%%%%%%%%%%%%%%%%%%%%%%%%%%%%%%%%%%%%%
%%%%%%%%%%%%%%%%%%%%%%%%%%%%%%%%%%%%%%%%%%%%%%%%%%%%%%%%%%%%%%%%%%%%%

\chap{The \nex2 superconformal structure of $W$-strings}

\sect{Notation and conventions}

We begin by introducing our notation and setting our conventions.  The
$W$-Liouville (or Toda) system will be realized by $n-1$ free bosons,
$\phi_{L,i}$, $i=1,2,\dots, n-1$, with operator product:
$\phi_{L,i}(z)\phi_{L,j}(w)=-\delta_{ij}\log(z-w)$.  It is natural to think of
them as generating
a Cartan subalgebra of $SL(n)$.  The energy momentum tensor is given by
$$
T_L\ =\ -\shalf \sum_{i=1}^{n-1}(\del\phi_{L,i})^2 +
\beta_0 (\rho\cdot \del^2\!\phi_L)\ , \EQN\TL
$$
where $\rho$ is the Weyl vector\foot{The Weyl vector is defined to be half the
sum of the positive roots.  We normalize roots to have squared-length equal to
two, and so for $SL(2)$ we take $\rho = 1/\sqrt{2}$.} of $SL(n)$,
$\beta_0\equiv{(t +1)\over\sqrt t}$, and $t$ is a parameter.

It is convenient to parametrize the central charge of the matter system as:
$$
c^M\ =\ (n-1)\Big[ 1-n (n+1) {(t-1)^2\over t}\Big]\ .\EQN\cmdef
$$
At this point we are not making any assumptions about the structure of the
matter system. It can be an arbitrary conformal field theory with an additional
$W_n$ symmetry.  At some points we will find it useful to realize the matter
system in terms of free bosons, $\phi_{M,i}$, $i=1,2,\dots n-1$, with
$\phi_{M,i}(z)\phi_{M,j}(w)=-\delta_{ij}\log(z-w)$ and with energy momentum
tensor:
$$
T_M\ =\ -\shalf \sum_{i=1}^{n-1}(\del\phi_{M,i})^2 +
i\alpha_0 (\rho\cdot \del^2\!\phi_M)\ , \EQN\TM
$$
where $\alpha_0\equiv{(1 - t)\over\sqrt t}$. Note that $W_n$-minimal matter
models, which we will denote by $\mpq^{(n)}$, correspond to
$t=q/p$. To get these minimal models, one must also perform a Felder \brs
reduction\foot{We will use the nomenclature ``Felder \brs reduction'' to
distinguish such a reduction from the \brs reduction using the \brs charge of
the underlying string theory.}  using the screening currents:
$$
S_M^\pm(\a_i) = e^{- i \a_\pm \alpha_i \cdot \phi_M} \ ,\EQN\matscreen
$$
where the $\alpha_i$'s are simple roots of $SL(n)$ and $\a_+\equiv \sqrt t$,
$\a_-\equiv -1/\sqrt t$  \FLZ.
In the free field  realization, the primary fields are represented by vertex
operators
$$
\Phi_{r_i;s_i}\ =\ V^M_{r_i;s_i}\ \equiv\ e^{i\a^M_{r_i;s_i}\cdot\phi_M}\  ,
\EQN\vertexm
$$
with
$$
\a^M_{r_i;s_i} \ = \ \sum_{j=1}^{n-1} [\a_+(r_j-1)+\a_-(s_j-1)]\,\l_j\
,\EQN\apm
$$
where $\l_i,\ i=1,\dots,n-1$, are the fundamental weights.
For minimal models $\mpq^{(n)}$, one can restrict the labels to the Kac table,
that is, to $\sum r_i \leq p-1$ and $\sum s_i \leq q-1$, $s_i\geq1, r_i\geq1$.
In addition, it will prove convenient to introduce vertex operators in the \lv\
sector
$$
V^L_{r_i;s_i}\ \equiv\ e^{\a^L_{r_i;s_i}\cdot\phi_L}\  ,
\EQN\vertexl
$$
where
$$
\a^L_{r_i;s_i} \ = \ \sum_{j=1}^{n-1} [-\a_+(r_j-1)+\a_-(s_j-1)]\,\l_j\
.\EQN\apl
$$
For the remainder of this section there will be no need to represent the matter
system by $n-1$ free bosons: it is only required that the matter system provide
a $W$-algebra and have central charge given by \cmdef.

The ghost system consists of ghosts, $c_j$, and anti-ghosts, $b_j$,
($j=1,2,\dots, n-1$) of spins $-j$ and $j+1$ respectively.  The corresponding
energy-momentum tensors are:
$$
T_{gh}^j(z) ~\equiv~ - (j+1) b_j \del c_j ~-~ j(\del b_j) c_j \ , \EQN\Tgh
$$
with central charge $c^{(j)} = -2(6j^2 + 6j +1)$. With the foregoing choices,
the total ghost central charge of $-2(n-1)(2n^2 + 2n +1)$ is precisely offset
by the central charge of the matter plus gravity systems: $c^L + c^M = $ $2n +
12(\beta_0^2 - \alpha_0^2) \rho^2 =$ $ 2(n-1)(2n^2 + 2n + 1)$.

The ``$W_n$-string'' models \multref\BG{\Das\LPSX}\MS\ that we will consider
are thus tensor products
$$
\cW^{(n)}\ \equiv\ W_n^{\rm matter}\otimes W_n^{\rm \lv}\otimes_{j=1}^{n-1}
\{b_j, c_j\} \ , \EQN\wstring
$$
where $\{b_j, c_j\}$ denotes the Hilbert space of the $j^{\rm th}$ ghost
system.  Since the \brs charges for these $W_n$-string systems are only
explicitly known for $n=2$ and \nex3, we now restrict to these cases, but we
will reconsider general values of $n$ in sections 3 and 5.

\sect{N=2 superconformal symmetry in two-dimensional gravity}

We start with non-critical $W_2$-strings, which are simply matter coupled to
ordinary gravity.  The matter system can be arbitrary, except that it must have
 conformal anomaly equal to $c^{M}=13-6(t+1/t)$.
The BRST current is well known \FMS:
$$
\jbrs(z) ~=~ c_1(z) \Big[ T_L(z) ~+~ T_M(z) ~+~ \shalf T_{gh}^1(z) \Big] \ .
\EQN\mgbrst
$$
It has been noted by many authors that the BRST current $J_{BRST}$ and the
antighost field $b_1(z)$ generates an algebra that is very similar to \nex2
superconformal algebra. Unfortunately this algebra is not closed. The operator
product of BRST current with itself is singular and generates a new field
$c\partial c$. However one can modify the BRST current by a total derivative
term in such a manner that the operator product is regular \BGS. Such a
modification clearly does not affect the BRST charge. To be explicit, we
introduce the $U(1)$ current:
$$
J(z) ~=~ c_1 b_1  ~+~ \sqrt{\coeff2t}~\del \phi_L \ , \EQN\mguone
$$
where $\phi_L$ is the Liouville field.  Then, defining an ``improved'' BRST
current,
$$
G^+(z) ~=~ \jbrs(z) ~-~ \sqrt{\coeff2t} \del (c_1 \del\phi_L) ~+~ \shalf
\big(1 - \coeff2t\big) \del^2 c_1 \ , \EQN\mggplus
$$
the currents $G^+(z)$ and $J(z)$, along with
$$
T(z) ~=~ T_L(z) + T_M(z) + T_{gh}^1(z) \qquad {\rm and} \qquad G^-(z) ~=~
b_1(z) \, \EQN\TG
$$
close into a twisted \nex2 superconformal algebra \TOPALG:
$$
\eqalign{
T(z)\shdot T(w)\ &\sim\
{2T(w)\over\zmw 2}
+{\partial T(w)\over (z-w)}\ ,\cr
T(z)\shdot G^\pm(w)\ &\sim\
{\coeff12(3\mp 1) G^\pm(w)\over\zmw 2}+{\partial G^\pm(w)\over (z-w)}\ ,\cr
T(z)\shdot J(w)\ &\sim\ {\coeff 13c^{N=2}\over\zmw 3} +
{J(w)\over\zmw 2}+{\partial J(w)\over (z-w)}\ ,\cr
J(z)\shdot J(w)\ &\sim\ {\coeff 13c^{N=2}\over\zmw 2}\ ,\ \ \ \
J(z)\shdot G^\pm(w)\ \sim\ \pm {G^\pm(w)\over (z-w)}\ ,\cr
G^+(z)\shdot G^-(w)\ &\sim\
{\coeff 13c^{N=2}\over\zmw 3}+{J(w)\over\zmw 2}+{T(w)+\partial J(w)
\over (z-w)}\ ,\cr
G^\pm(z)\shdot G^\pm(w)\ &\sim\ 0\ .\cr
}\EQN\tneqtwo
$$
The anomaly, which is proportional to the central charge of the corresponding
untwisted \nex2 superconformal theory, is:
$$
c^{N=2} ~=~ 3\big(1 ~-~ {2 \over t}\big) \ . \EQN\minntwo
$$
This relation implies that the string with $c^M=1$ has $c^{N=2}=-3$, while the
ordinary bosonic string (with $c^M=25$) is mapped onto a superconformal theory
with $c^{N=2}=9$.  An immediate observation \BGS\ is that if
the matter model is taken to be $\mm{p=1}{q=k+2}^{(2)}$, then this central
charge becomes $c^{N=2} = {3k \over (k+2)}$, and corresponds to the \nex2
superconformal minimal models. We will discuss this point in more detail in
section 4.

One can also obtain a twisted \nex2 superconformal algebra by replacing $t \to
1/t$ throughout the foregoing discussion.  Therefore the general matter-gravity
system has at least two  twisted \nex2 superconformal algebras.

It is interesting to note that the foregoing construction of the \nex2
superconformal operators did not require any specific form for the matter part
of the theory. The only requirement on the combined matter and gravity systems
is the existence the $U(1)$ current that one uses to modify the BRST current.
The matter system can be arbitrary and thus even the critical ($D = 26$) string
theory possesses \nex2 twisted supersymmetry. If one can find a realization of
the matter sector that employs at least one boson, $\phi_M$, then one can use
this, instead of $\phi_L$, to modify the \brs current \foot{This possibility is
utilized in a forthcoming paper by C.Vafa and S.Mukhi to relate the $c^M=1$
string to a superconformal $SL(2)/U(1)$ coset model \ref\vmuk{C. Vafa and S.
Mukhi, private communication.}.}. This choice is equivalent to simply
performing the substitution $t \to -t$ and interchanging $\phi_L$ with $\phi_M$
throughout this section.

Having exposed this \nex2 superconformal structure, it is natural to recast
general questions about gravity coupled to matter in terms of \nex2
superconformal theory.  We will defer this until section 4.

The appearence of an extended superconformal structure appears to be
quite general in systems of matter with extended symmetries, coupled to
extended geometries.

\sect{N=3 superconformal structure of N=1 supergravity}

In this section we will generalize the construction discussed above to
superstrings. It was suggested by several authors that topological superstring
theories may have \nex3 twisted supersymmetry \ref\fuj{K.~Fujikawa and
H.~Suzuki, \nup{361} (1991) 539; H.~Yoshii, ``Hidden Higher Supersymmetries in
Topological Conformal Field Theories,'' Tokyo University preprint TIT/HEP-165,
December 1990}. We will see below that any superstring theory has an \nex3
twisted supersymmetry.
The only requirement that we will make is that the
combined super-matter and supergravity system possess a $U(1)$ supercurrent
that can be used to improve the \brs current. We parametrize the matter
central charge as \eqn\anom{\hat c =5-2(t+1/t)~,} and in this parametrization
the superconformal minimal models correspond to rational $t=q/p$.

The modified BRST current $J_{BRST}$ (spin $1$) and the diffeomorphism ghost
$b$ (spin $2$) together with \nex1 supercurrent $G$ (spin $3/2$) constitute
three super-generators of \nex3 twisted superalgebra. As in the bosonic case,
the stress-energy tensor $T$, modified BRST current $J_{BRST}$, the ghost $b$
and the modified ghost current $H$ make an \nex2 multiplet. The original \nex1
supersymmetry remains unbroken so that we can use \nex1 superformalism
\eqn\super{\eqalign{{\cal J}_{BRST}&=j_{BRST}+ \theta J_{BRST} \ ,\cr
                     {\cal T}&=G+ \theta T\ ,\cr
                       {\cal B}&=\beta+\theta b \ ,\cr
                        {\cal H}&=h+\theta H \ .\cr}}
The lower components of the superfield make another \nex2 multiplet.
For later convenience we also introduce the super-ghost field
${\cal C}=c+\theta \gamma$.
The lower component of the BRST current $j_{BRST}$, modified $U(1)$
charge and the super-diffeomorphism ghost $\beta$ become the $SL(2)$
currents of \nex3 twisted algebra.

In order to identify $J_{BRST}$ with the supercurrent and
$j_{BRST}$ with one of the $SL(2)$ currents, we will modify the super-BRST
current
by total derivative piece
%${\cal J}_{BRST} \rightarrow  {\cal J}_{BRST}+ \Delta {\cal J}_{BRST}$
to ensure the operator products  $G^+(z)G^+(w)$  and $K^+(z)K^+(w)$
are regular. Namely
\eqn\sBRST{\eqalign{K^+ + & \theta G^+={\cal J}_{BRST}+ \Delta {\cal
J}_{BRST}\cr
{\cal J}_{BRST}&=({\cal T}+{1 \over 4}D{\cal C}D{\cal B}-
{3 \over 4}\partial{\cal C}{\cal B}){\cal C}-{3 \over 4} D({\cal C}D{\cal
C}{\cal B})\cr
\Delta {\cal J}_{BRST}&= D({\cal C}D{\cal C}{\cal B})-{1 \over 2t}D\partial
{\cal C}-
{1 \over \sqrt{t}}D({\cal C}\partial \Phi)+ {1 \over 2\sqrt{t}}D(D{\cal
C}D\Phi)~,\cr}}
where $D=\partial_{\theta}+\theta \partial_z$ and $\Phi=\phi + \theta \psi$ is
the  super-Liouville field.

We are not going to describe these rather tedious calculations in detail, but
we will just present the results.
The three $SL(2)$ currents are given as follows
\eqn\su{\eqalign{K^+ &=2i( j_{BRST}-{1 \over 2t}\partial \gamma -{1 \over
2\sqrt{t}}\gamma \partial\phi+
{1 \over 2}\partial c \psi+{1 \over \sqrt{t}}c\partial \psi)~,\cr
K^3 &=cb-\beta\gamma+{1 \over \sqrt{t}}\partial \phi~~,~~~~K^-=2i\beta \cr }}
It is straightforward to check that these currents make an honest $SL(2)$
current algebra at level $-1/t$. There are also three super-generators
\eqn\super{\eqalign{G^+ &=2i(J_{BRST}-{1 \over 2t}\partial ^2 c-
{1 \over \sqrt{t}}\partial (c \partial \phi)+
{1\over 2 \sqrt{t}}\partial (\gamma \psi) +\partial (c \beta \gamma))~,\cr
G^3&=2(G_{m+l}+G_{ghost})~~,~~~~G^-=-2i b \cr}}
and there is also an additional free fermion
$F=\beta c-{1 \over \sqrt{t}}\psi $.

Undoing the twisting: $T \rightarrow T-{1 \over 2}\partial K^3$ and $G^3
\rightarrow G^3+\partial F$, one obtains the \nex3 \sc algebra, described by
Seiberg and Schwimmer \ssh. All fields become conformal primary fields. The
supercurrents $G^a$ are almost in the adjoint representation of $SL(2)$
\eqn\adj{K^a G^b \sim {1 \over (z-w)}f^{ab}_c G^c - F {\eta^{ab} \over
(z-w)^2}~,}
modulo the fact that there is a central term proportional to the free fermion.
In the operator products between supercurrents one generates the stress-energy
tensor and $SL(2)$ currents
\eqn\opesuper{G^a G^b \sim {1 \over (z-w)^2}f^{ab}_c K^c +
{1 \over (z-w)}(2\eta^{ab} T +2f^{ab}_c \partial K^c)-
{2/t \over (z-w)^3}\eta^{ab}}
In the operator products between supercurrents and free fermion one generates
the $SL(2)$ currents
\eqn\name{G^a F \sim -{1 \over (z-w)}K^a}
The remaining operator products are
\eqn\ope{FF \sim {-1/t \over (z-w)}~{\rm and}~K^a F \sim Reg}
The resulting \nex3 algebra has a conformal anomaly equal to $c^{N=3} =-3/t$.
As
in section 2 of this paper, one can can replace $t \to 1/t$ in the foregoing
calculation and thereby obtain a theory with $c^{N=3} =-3t$. In the derivation
above we did not make any assumption about the structure of \nex1 matter
system. That means that even ten-dimensional critical superstring theory has
\nex3 twisted supersymmetry.

The discussion above implies that superconformal matter systems coupled to
supergravity are closely related to topological $N=3$ twisted models.
In order to construct these models explicitly one has to know the
representation theory  of $N=3$ superconformal algebra. We postpone the
discussion of the representation theory of \nex3 algebra to
one of our future publications.

It is a little puzzling to have an \nex3 algebra. We know quite well that for
$\sigma$-models, \nex3 supersymmetry automatically implies \nex4 \ref\alv{
L.~Alvarez-Gaume and D.Z.~Freedman, \cmp{80} (1981) 443.}. In fact, in our case
there are some reasons to believe that the \nex3 algebra discussed above
can be promoted to \nex4. This question is also under investigation.

\sect{N=2 superconformal structure of $W_3$-gravity}

A \brs charge for the $W_3$-string, $\cW^{(3)}$, was found in
\doubref\BLNW\BSS. This charge is nilpotent if the total central charge
vanishes ({\it i.e.}, if $c^L+c^M=-c^{gh}=100$), and it is given by
$\qbrs = \oint\!{dz\over2\pi i} \jbrs(z)$, where
$$
\eqalign{
\jbrs\ &=\ \ci2 \big[\coeff1{b_{L}}W_L  + \coeff i{b_{M}} W_M \big]
+ \ci1 \big[T_L +T_M +\shalf\tg^{1} +\tg^{2} \big]\cr
&+ \big[T_L -T_M \big]\bi1 \ci2(\del\ci2)
+ \mu(\del\bi1)\ci2(\del^2\!\ci2)+ \nu\bi1 \ci2(\del^3\!\ci2)\ ,}\EQN\jb
$$
and where $b^2_{L,M}\equiv{16\over5c_{L,M}+22}$
and $\mu={3 \over 5}\nu= {1 \over
10{b_L}^2} (1-17{b_L}^2)$. In this equation, $T_{L,M}$ and $W_{L,M}$ denote the
usual stress tensors and $W$-generators of the \lv\ and matter sectors, and
$\tg^i$ are the stress tensors of the ghosts.

The \brs current, $\jbrs$, can be identified with the \nex2 supercurrent,
$G^+$, if one adds a total derivative piece to $\jbrs$ so as to ensure that
$G^+(z)G^+(w)\sim 0$. That is, for
$$
G^+\ =\ \jbrs + \Delta\cJ \EQN\improv
$$
we find:
$$
\eqalign{ \Delta\cJ\ &=\
\del\Big[\,
-c_{1}J+2i\sqrt{\coeff t3} b_{1} c_{1} c_{2}J +
  i\coeff{\left( 1 + t \right) }{2} \sqrt{\coeff3t}
   b_{1} c_{1} (\del c_{2})
   \cr&-
    i\coeff{\left( 3 + 2 t \right) }{\sqrt{3 t}}
   b_{1} (\del c_{1}) c_{2} -
    \coeff{(7 {t^2}-10t-15)}{4 t}   b_{1} (\del^{2}\!c_{2}) c_{2}+
  i\coeff{(t-9)}{\sqrt{3 t}} b_{2} (\del c_{2}) c_{2}
  \cr&-
  i\coeff{\left( 3 + 4 t \right) }{\sqrt{3 t}} (\del b_{1}) c_{1} c_{2}-
  \coeff{3 (4 {t^2}-2t-3)}{2 t} (\del b_{1}) (\del c_{2}) c_{2} +
 \coeff{(t-3)}{t} (\del c_{1})
  \cr&+
 i\coeff{1}{2 \sqrt{3 t}} c_{2}[2t {J^2}-3 (t-5)T_L-3 (t-1)T_M
 -6( 1 + t )\del J]
  \cr&+
  i\coeff{ \left( 1 + t \right) }{2} \sqrt{\coeff3t}(\del c_{2})J-
  i\coeff{({t^2}-4t-1)}{2t}\sqrt{\coeff3t} (\del^{2}\!c_{2}) +
  t b_{1} (\del c_{2}) c_{2}J
  \, \Big]\ .}\EQN\deltaj
 $$
Here, $J$ in \deltaj\ denotes the $U(1)$ current:
$$
J\ =\  c_{1} b_{1}  + c_{2} b_{2}  +
  \coeff3{\sqrt t}(\l_1\cdot \del \phi_L)
 +\coeff i2\sqrt{\coeff3t}  (t-1) \del[b_{1}c_{2}]\ ,\EQN\jone
$$
where $\l_1$ denotes the first fundamental weight
of $SU(3)$.  In deriving  \deltaj\ there is a choice of sign, and the other
choice leads to the conjugate fundamental weight, $\l_2$, in \jone.

It is straightforward, but rather tedious, to check that
$J$, $G^+, G^-\equiv b_1$, and
$T\equiv T_M+T_L+\tg^{1} +\tg^{2}$ generate a (topologically twisted) \nex2
superconformal algebra \tneqtwo, with anomaly
$$
c^{N=2} ~=~ 6\big(1 ~-~ {3 \over t}\big) \ . \EQN\cks
$$
It is analogous to the \nex2 superconformal algebra that we found for matter
models coupled to ordinary gravity. However, since we are now coupling models
with extended symmmetry to $W$-gravity, it is natural to expect an extension of
the \nex2 superconformal algebra. One may indeed check that the \brs multiplet
of currents
$$
\eqalign{
V_2&\ =\ \alpha\Big[
  \left( 2 + t \right)(3-t)T_M -   ({t^2}+4t-24)T_L+
i\coeff{3t }{{\sqrt{3t}}}(5 t-18) T_L b_{1} c_{2}
\cr&-
  \coeff{{t} }{6} (t-12)J^2 +
  i\coeff{{t^{2}} (5 t-18) }{3 {\sqrt{3t}}}J^2 b_{1} c_{2} -
  \coeff{  {t} (5 t-18)}{3} J b_{1} c_{1}-
  i\coeff{{t} (t-3) (5 t-18)}{6 {\sqrt{3t}}}J b_{1} (\del c_{2})
   \cr&+
 i\coeff{2 {t^2} (5 t-18)}{3
    {\sqrt{3t}}} J (\del b_{1}) c_{2} -
  \coeff{2   {t} (5 t-18)}{3} b_{1} b_{2} c_{1} c_{2} +
  i\coeff{{t} \left( 9 + 5 t \right)  (5 t-18)}{3
    {\sqrt{3t}}} b_{1} b_{2} (\del c_{2}) c_{2}
    \cr&-
  i\coeff{t \left( 1 + t \right)  (5 t-18)}{{\sqrt{3t}}}
   b_{1} c_{2} (\del J) + \coeff{(11 {t^2}-24t-36)}{3}
   b_{1} (\del c_{1}) -i\coeff{ (5 t-18)
    ({t^2}-6t-9)}{3 {\sqrt{3t}}} b_{1} (\del^{2}\!c_{2})
    \cr&+
  i\coeff{{\sqrt{3t}}(5 t-18)}{3} b_{2} c_{1}+
  \coeff12  (11 {t^2}-9t-90) b_{2} (\del c_{2}) +
  \coeff16(31 {t^2}-81t-90) (\del b_{1}) c_{1}
  \cr&+
  i\coeff{{t}\left( 3 + 7 t \right)  (5 t-18) }{3{\sqrt{3t}}}
     (\del b_{1}) b_{1} c_{1} c_{2} +
  \coeff12\left( 1 + t \right) (3 t-5) (5 t-18)
   (\del b_{1}) b_{1} (\del c_{2}) c_{2}
   \cr&-
  i\coeff1{12{\sqrt{3t}}}
    (5 t-18) (23 {t^2}-6t-45) (\del b_{1}) (\del c_{2}) +
  \coeff13( 16 {t^2}-27t-90) (\del b_{2}) c_{2}
  \cr&-
  i\coeff1{6 {\sqrt{3t}}}(5 t-18) (5 {t^2}-9)
   (\del^{2}\!b_{1}) c_{2}
  +\coeff16 ( 13 {t^2}-72-24t) (\del J)\, \Big]
 \cr
F^+_2&\ =\ - [\qbrs,V_2] \cr
F^-_3&\ =\ \ [(G^-)_{-1},V_2] = \alpha\Big[
-\coeff23 {t} (3-2t) b_{1} J +
  \coeff23{t} (5 t-18) b_{1} b_{2} c_{2}
 \cr &\qquad\qquad\ \  +
  i\coeff{{t^2} (5 t-18)}{{\sqrt{3t}}}
   (\del b_{1}) b_{1} c_{2} -
 i\coeff{t(5 t-18)}{\sqrt{3t}} b_{2} -
  \coeff32(3 {t^2}-9t-2)(\del b_{1})\Big]
 \cr
W_3&\ =\ \shalf\{\qbrs,F^-_3\}
}\EQN\superw
$$
forms a $W_3$ extension \doubref\NY\larry\ of the (topologically twisted) \nex2
superconformal algebra. The currents are correctly normalized, {\it i.e.},
$V_2(z)V_2(w)\sim {{1\over2}c^{N=2}\over (z-w)^4}+\dots$, if one sets
$\a=[(t-4)(t+2)(2t-3)(5t-18)/3]^{-1/2}$.
Note that this \nex2 super-$W_3$ algebra exists for
generic\foot{At special values of $t$, the algebra may truncate. For instance,
for $t=4$, the $V$ multiplet becomes null and the algebra truncates
to the \nex2 superconformal algebra.} $t$. Note also that the matter system
enters only via its $W$ currents and therefore the foregoing statements apply
to any $W_3$ matter system.

For the special minimal $W$-matter models $\cM_{1,q}^{(3)}$, $t$ is integer and
writing $t=q=k+3$, the central charge in \cks\ becomes identical to the central
charge of the \nex2 superconformal coset models, or Kazama-Suzuki models \KS,
based on cosets $SU(3)_k/U(2)$. This is natural in view of our conjecture
\BLNW\ that for general $n$, the models $\cM_{1,q=n+k}^{(n)}$ coupled to
$W_n$-gravity should be closely related to (or even be equivalent to)
topologically twisted \nex2 superconformal coset models based on
$SU(n)_k/U(n-1)$, coupled to topological $W_n$-gravity. It is indeed well-known
\refs{\ITO{,}\NY{,}\larry{,}\LNW}\ that the Kazama-Suzuki model based on
$SU(n)_k/U(n-1)$ has an \nex2 super-$W_n$ algebra as its maximal chiral
algebras, and has a central charge
$$
c^{N=2}\ =\ 3{(n-1)k\over n+k}\ . \EQN\cnk
$$
One thus suspects that, in general,
$W_n$-matter models coupled to $W_n$-gravity should exhibit an \nex2
superconformal $W_n$ symmetry, with central charge \cnk.
That is, besides the spin one supermultiplet, or \brs multiplet,
$(J,G^+,G^-,T)$, there should also be, for each spin $s$, a supermultiplet of
of currents of the form
$$
\cV_s\ =\ \left\{\matrix{
V_s \cr F^+_{s\phantom{+1}}= -[\qbrs,V_s] \cr F^-_{s+1}\,= \ [(G^-)_{-1},V_s]
\cr W_{s+1}\ \,=\shalf\{\qbrs,F^-_{s+1}\}}
\right\}\ \ \qquad \ s=2,\dots,n-1\ .\EQN\wcurr
$$
The currents in this multiplet are composed of
matter and \lv\ $W$-currents, the
$U(1)$ current  and the ghosts.  In the next section we will give an explicit
construction of such ``non-standard'' \nex2 super-$W_3$ algebras and we will
also establish the presence of \nex2 superconformal $W_n$ symmetry in general.

It appears that \nex2 superconformal extensions of $W$-algebras carry some
intrinsic notion of $W$-geometry. In particular, one can immediately infer the
full structure of the descent equations of $W_n$-gravity. For example, for
$W_3$ one can associate with each \brs\ invariant operator $\Phi^{(0)}$ fields
at adjacent ghost numbers in the following manner (the superscripts label
relative ghost number):
$$
\eqalign{
\Phi^{(-1)}_{1}\ &=\ (G^-)_{-1}\Phi^{(0)} \cr
\Phi^{(-1)}_{2}\ &=\ (F^-_3)_{-2}\Phi^{(0)}\ \sim\ [(b_2)_{-2}+\dots]\Phi^{(0)}
\cr
\Phi^{(-2)}\ &=\ (F^-_3)_{-2}(G^-)_{-1}\Phi^{(0)}\ \sim\
[(b_2)_{-2}(b_1)_{-1}+\dots]\Phi^{(0)} \cr
\Phi^{(-3)}\ &=\ (F^-_3)_{-1}(F^-_3)_{-2}(G^-)_{-1}\Phi^{(0)}\ \sim\
[(b_2)_{-1}(b_2)_{-2}(b_1)_{-1}+\dots]\Phi^{(0)} \cr
& \qquad \dots, etc.}\EQN\avatars
$$
{}From the explicit form \doubref\NY\larry\ of the \nex2 $W_3$ algebra it is
then clear that these fields satisfy
$$
\eqalign{
[\qbrs,\Phi^{(0)}]\ &=\ 0  \cr
\{\qbrs,\Phi^{(-1)}_{1}\}\ &=\ L_{-1}\Phi^{(0)}  \cr
\{\qbrs,\Phi^{(-1)}_{2}\}\ &=\ 2\{(W_3)_{-2}+(V_2)_{-2}\}\Phi^{(0)}  \cr
[\qbrs,\Phi^{(-2)}]\ &=\
2\{(W_3)_{(-2)}+(V_2)_{-2}\}\Phi^{(-1)}_{1}-L_{-1}\Phi^{(-1)}_{2}\cr
\{\qbrs,\Phi^{(-3)}\}\ &=\
\{2(W_3)_{-1}+(V_2)_{-1}\}\Phi^{(-2)}-(F^-_3)_{-1}[\qbrs,\Phi^{(-2)}]\cr
& \qquad  \dots, etc.}\EQN\descent
$$
Such descent equations generalize those of topological gravity
\doubref\Witopgr\VVtopgr, and will be important for defining integrals over
$W$-moduli space \BLNW\ in correlation functions.

%%%%%%%%%%%%%%%%%%%%%%%%%%%%%%%%%%%%%%%%%%%%%%%%%%%%%%%%%%%%%%%%%%%%%
%%%%%%%%%%%%%%%%%%%%%%%%%%%%%%%%%%%%%%%%%%%%%%%%%%%%%%%%%%%%%%%%%%%%%
%%%%%%%%%%%%%%%%%%%%%%%%%%%%%%%%%%%%%%%%%%%%%%%%%%%%%%%%%%%%%%%%%%%%%

\chap{Hamiltonian reduction}

In the last section we presented several examples of twisted \nex2
superconformal algebras. Somewhat surprisingly, the free field realization of
the \nex2 superconformal algebra was not of the standard form
\refs{\Muss{,}\ITO{,}\NY}. For example, for the minimal models we found a free
field realization that contained two free bosons and two fermionic fields, but
the latter had conformal dimension ${3 \over 2}$ and $-\half$. In this section
we show how different free field realizations of the \nex2 superconformal
algebra, and its $W$-extensions, can be obtained by making different choices of
the Borel subalgebra in a hamiltonian reduction of the superalgebra
$SL(n|n-1)$. We present only the classical arguments, but from our analysis it
is clear how one can generalize to the quantum case. From our discussion it
will also be clear that two-dimensional matter coupled to $W$-gravity is
naturally embedded in a constrained Wess-Zumino-Witten model based on
$SL(n|n-1)$.

It is well known that the hamiltonian reduction of the super-algebra
$SL(n|n-1)$ gives rise to the $W_n$-extension of the \nex2 superconformal
algebra \doubref\ITO\NY. To carry out the reduction it is convenient to use the
Gauss decomposition ${\cal N}_- \oplus{\cal H} \oplus {\cal N_+}$ of the
algebra. In this decomposition one imposes constraints on the currents in
${\cal N_+}$, and the remaining degrees of freedom (${\cal N_-}$) are gauged
away. It is not surprising that different choices of Gauss decomposition can
give rise to different representations of the same chiral algebra. The free
field realization of the chiral algebra can be deduced from the Wakimoto
realization of the original current algebra. Each choice of Borel subalgebra
will give rise to a different Wakimoto representation of the $SL(n|n-1)$
current algebra and hence to a different representation of the same extended
\nex2 superconformal algebra. Below we compare two different schemes of
reducing the superalgebra $SL(n|n-1)$. There does not exist a simple way of
going from one realization to the other one. The two different free field
realizations are related to each other by some highly non-trivial Bogolubov
transformation.

To illustrate the basic idea of different reductions, we start by considering
the simple example of $SL(2|1)$.   The super-algebra $SL(2|1)$  can be realized
in terms of  $3 \times 3$ matrices with the following  assignments
of statistics
\eqn\sltwo{ \bordermatrix {& B & F & B \cr B & \ & j_{\alpha_1} & J_{\alpha_1+
\alpha_2} \cr
F & j_{-\alpha_1}&  \ & j_{\alpha_2} \cr
B &  J_{-(\alpha_1 + \alpha_2 )} & j_{-\alpha_1} &  \  \cr} \ . }
The nilpotent subalgebra ${\cal N_+}$ consists of the upper triangular matrices
and  is generated by two
fermionic currents $j_{\alpha_1},~j_{\alpha_2}$
\eqn\nil{\{j_{\alpha_1}, j_{\alpha_2}\}=J_{\alpha_1+\alpha_2}}

The  \nex2 superconformal algebra is obtained  by  constraining
 the $SL(2|1)$ currents in the Borel sub-algebra
\eqn\const{j_{\alpha_1}=\chi^{\dagger}~~,~~~j_{\alpha_2}=\chi~~{\rm
and}~~~J_{\alpha_1+\alpha_2}=1~,
}
In order to make the constraints  \const\ first class we  have introduced a
pair of conjugate fields $\chi(z)$ and $ \chi^{\dagger }(z)$ of conformal
dimension ${1 \over 2}$, with operator product:
\eqn\auxf{ \chi(z) \chi^{\dagger}(w) \ = \ { 1 \over z-w } \ .}
We will not discuss in detail how the reduction works for this case, but we
refer the reader to the original papers on hamiltonian reduction \bo. The free
field realization can be obtained from a generalized Miura transformation.
Consider the operator $L=\partial_z - {\cal J}(z)$, where ${\cal J}(z)$ is a
constraint current. This operator naturally appears in the context of
constrained Wess-Zumino-Witten models. The constrained Wess-Zumino-Witten model
is invariant under gauge transformations generated by lower triangular
matrices. There are two natural choices for the gauge slice. First choose the
current ${\cal J}$ in such a way that it is consistent with the constraints
\const
\eqn\free{ {\cal J}(z)  \ = \ \left( \matrix{{i \over \sqrt{2}} (\del \phi_1 +
\del \phi_2 ) & \chi & 1 \cr
0 & i\sqrt 2 \del \phi_2 & \chi^{\dagger } \cr
0 &0 & {i \over \sqrt 2 }(-\del \phi_1 + \del \phi_2 ) \cr } \right ) \ \ .}
The second choice corresponds to writing ${\cal J}(z)$ in terms of the fields
in the chiral algebra.  If we perform a gauge transformation
 by lower triangular matrices  we  can reduce \free\ to the following form
\eqn\super{{\cal J}(z) \ = \ \left ( \matrix{ J&0& 1 \cr G^- & 2J & 0 \cr
T-J^2&  G^+ &J\cr } \right )  \ \ . }
It is easy to see that fields $J, G^{\pm}$ and $T$ form an \nex2 superconformal
algebra. Simple calculations lead to the following well known bosonisation
rules \Muss\ for the \nex2 superconformal algebra:
\eqn\bos{\eqalign{J \ & = \ - \half \chi^{\dagger } \chi + {i \over \sqrt 2
}\del \phi_2 \cr
G^{-} \ & = \ {i \over \sqrt 2} \chi^{\dagger }( \del\phi_2 - \del\phi_1) -
\del \chi^{\dagger } \cr
G^{+} \ & = \ {i \over \sqrt 2 }\chi   (\del\phi_2 + \del\phi_1) + \del \chi
\cr
T\ & = \  -\half (\del \phi_1)^2 -\half (\del \phi_2)^2 -\half \chi^{\dagger}
\del \chi -\half \chi \del \chi^{\dagger} +{i \sqrt 2} \del^2 \phi_1 \ \ .
\cr}}
The screening currents can also be obtained from the hamiltonian reduction. In
the free field realization of the superalgebra $SL(2|1)$ we have a screening
operator for every simple root generator of Borel sub-algebra. After one
imposes the constraints, these screening currents turn into screening currents
of the \nex2 superconformal algebra
\eqn\nescreen{S_{\alpha_1} = \chi e^{i\alpha_- (\phi_1 + i \phi_2)}
\quad { \rm and } \quad   S_{\alpha_2} = \chi^\dagger e^{i \alpha_- ( \phi_1 -
i \phi_2)} \ . }

The  representation of \nex2 superconformal algebra encountered in the last
section ({\it i.e}., that given in in \mguone--\TG) is obtained by choosing
a different Gauss decomposition. As before we realize $SL(2|1)$ in terms of
$3 \times 3$ matrices  but with the following assigments of statistics:
\eqn\sltown{\bordermatrix{ & B & B & F \cr B & \ & J_{\alpha_1 +\alpha_2}
& j_{\alpha_1} \cr B & J_{-(\alpha_1 +\alpha_2 )} & \ & j_{-\alpha_2} \cr
F& j_{-\alpha_1} & j_{\alpha_2} & \  \cr
}}
Here we have chosen our matrix in such away that a the first two rows and
columns are  bosonic.
The Borel subalgebra is given by upper triangular matrices.
These different choices of the Borel subalgebra are related to each other by
a Weyl reflection with respect to a  fermionic root.
In this realization one of the simple roots is bosonic
and  the other one is fermionic. The nilpotent subalgebra ${\tilde {\cal N}}_+$
is generated by $J_{\alpha_1 +\alpha_2}$ and $j_{-\alpha_2}$
\eqn\nill{[J_{\alpha_1 +\alpha_2}, j_{-\alpha_2}]=j_{\alpha_1} \ . }
General arguments imply that different reduction  schemes with
respect to ${\cal N}_+$  generated by $j_{\alpha_1},~j_{\alpha_2}$ and
$J_{\alpha_1+\alpha_2}$
(see \nil) and ${\tilde {\cal N}}_+$  give rise to the same \nex2
superconformal theories \ref\KPpc{ K.~Pilch, private communication.}.
As before we impose  constraints on the currents in the nilpotent subalgebra
${\tilde {\cal N}}_+$
\eqn\fircons{ j_{\alpha_1} =0~~,~~~j_{-\alpha_2 }= \lambda^{\dagger }~~
{\rm and}~~~J_{\alpha_1 + \alpha_2}=1 \ ,}
where $ \lambda^{\dagger } $ is an auxiliary field with conformal dimension $ 3
\over 2$. Surprisingly the field conjugate to $\lambda^{\dagger }$ does not
appear in the constraints. To identify the conjugate field, consider how ${\cal
J}(z)$ transforms under a gauge transformation. As before the theory is gauge
invariant under ${\cal J}(z) \rightarrow {\tilde{\cal J}}(z)=S\partial_z S^{-1}
-SJ(z)S^{-1}$, where $S$ is lower triangular matrix. The field conjugate to
$\lambda^{\dagger }$ can be identified with one of the entries in matrix $S$
\eqn\ss{S \ = \ \left ( \matrix { 1 & 0 & 0 \cr a & 1 & 0 \cr b & \lambda & 1
\cr } \right ) \ . }
 This is nothing other than an  additional gauge fixing condition
that naturally appears in the context of constrained Wess-Zumino-Witten models.
As before, the bosonisation rules are obtained by  comparing  two different
gauge slices. From the constraints \fircons\ we have
\eqn\freegauge{{ \cal J}(z) \ = \ \left (\matrix {{i \over \sqrt 2}(\del \phi_1
+ \del \phi_2) & 1 & 0 \cr
0 &{i \over \sqrt 2}(- \del \phi_1+ \del \phi_2) & \lambda^{\dagger } \cr
0&0& i \sqrt 2\del \phi_2 \cr} \right )  \ \ . }
By choosing an appropriate gauge transformation one can always reduce
$\tilde{\cal  J}(z)$ to the following form
\eqn\susy{\tilde{\cal J}(z) \ = \ \left ( \matrix {-J & 1& 0 \cr
T-J^2 & -J & G^{-} \cr
G^{+} & 0 & -2J \cr}  \right )  \ .}
One can easily convience oneself that fields $J, G^{\pm}$ and $T$ generate an
\nex2 superconformal algebra.
The  bosonisation rules read\foot{These bosonisation rules have also been
obtained by Lev Rozansky \ref\lev{L. Rozansky, private communicaton}.}
\eqn\newbos{\eqalign{J \ & =\ \half \lambda \lambda^{\dagger} -
{i \over \sqrt{2}} \del \phi_2 \cr
G^{-} \ & = \ \lambda^{\dagger} \cr
G^{+} \ &  = \ \half \lambda \lambda^{\dagger}\del \lambda - {\lambda\over 2 }
((\del \phi_1)^2 + (\del \phi_2)^2)\cr
T(z) \ & = - \half (\del \phi_1)^2 - \half (\del\phi_2)^2 - {3 \over 2 }
\lambda^{\dagger} \del \lambda  +
 \half \lambda \del\lambda^{\dagger}  - {i \over \sqrt 2 } \del^2\phi_1
\ . \cr }}
These classical formulas can be easily generalized to the quantum case.
hamiltonian reduction also allows us to construct the screening currents.
Reducing the screening currents of $SL(2|1)$ we obtain
\eqn\screen{ S_{\alpha_1 + \alpha_2} = e^{2i \alpha_- \phi_1 }
{}~~{\rm and}~~~ S_{-\alpha_2} = \lambda^{\dagger}
    e^{- i\alpha_-(\phi_1+ i \phi_2 )}}

Comparing with \mguone--\TG\ we can immediately identify $\lambda$ with the
diffeomorphism ghost $c$ and $\lambda^{\dagger}$ with the anti-ghost $b$. In
addition, $\phi_1$ can be identified with the matter free field, $\phi_M$, and
$\phi_2$ can be identified with the \lv\ field, $\phi_L$. Note that in order to
make these identifications, one also has to twist the stress-energy tensor. In
passing from the classical result to the quantum version, one finds that the
improvement terms of \mggplus\ appear as quantum corrections to \newbos.

{}From the analysis above, it is clear how to generalize this construction to
the superalgebra $SL(n|n-1)$. The standard representation of the super-$W_n$
algebra is obtained by choosing a Borel sub-algebra generated by the fermionic
simple roots $ \alpha_1, \ldots, \alpha_{2n-2}$ of $SL(n|n-1)$. We can
represent $SL(n|n-1)$ in terms of $(2n-1) \times (2n-1)$ matrices and assign
the statistics to the entries in the chess order as in \sltwo, with alternating
bosonic and fermionic labels. It is known that if one imposes the following
constraints:
\eqn\newconst{\eqalign{j^+_{\alpha_1}&=\chi_1~~,
{}~~~j^+_{\alpha_2}=\chi_2+\chi^{\dagger}_1~~
,\ldots~j^+_{\alpha_{2n-2}}=\chi^{\dagger}_{2n-3}\cr
J^+_{\alpha_1+\alpha_2}&=1~~,~~~J^+_{\alpha_2+\alpha_3}=1~~,
\ldots~J^+_{\alpha_{2n-3}+\alpha_{2n-2}}=1\cr}}
and sets the rest of the currents in the upper triangular subalgebra to zero,
then the reduced theory has
a super-$W_n$ algebra \doubref\ITO\NY.
The whole discussion is completely parallel to that for $SL(2|1)$, and we will
not describe it here.

The other representation of super-$W_n$ algebra utilizes a different Gauss
decomposition of the superalgebra $SL(n|n-1)$. Again we realize
$SL(n|n-1)$ in terms $(2n-1) \times (2n-1)$ matrices and but assign statistics
in such a way that the first  $n$ rows and
$n$ columns are  bosonic.  The nilpotent subalgebra is generated by
$2n-3$ bosonic simple roots and one fermionic simple root. The constraints are
quite complicated for the general case.
First, we  consider $SL(3|2)$, whose constrained form is:
\eqn\example{\bordermatrix{ &B&B&B&F&F \cr
B&   \star & 1  & 0 &
0 & 0 \cr
B& \star  &   \star & 1 & 0 & 0 \cr
B & \star &  \star &    \star & \lambda^{\dagger}_1 & \lambda^{\dagger}_2 \cr
F& \star  & \star & \star &   \star & 1+ \lambda_1 \lambda^{\dagger}_2 \cr
F &  \star  & \star  & \star  & \star  &    \star \cr }~. }
We have introduced two sets of auxiliary fields
$(\lambda_1,~\lambda^{\dagger}_1)$  with conformal dimension  $(-{3 \over 2},{5
\over 2})$ and $(\lambda_2,~\lambda^{\dagger}_2)$ with conformal   dimension
$(-{1 \over 2},{3 \over 2})$.
As before the field, $\lambda_2$, that is conjugate to $\lambda_2^\dagger$
does not appear in the constraints. The field $\lambda_2$  has to identified
with one of the parameters of the  gauge transformation.
To get the free field realization of super-$W_3$ we have to write the relation
between different gauge slices. The first gauge slice is obtained by
constraining all the currents from ${\cal N}_-$ to zero (as in \freegauge).
The second gauge slice is given by:
\eqn\sltgau{\left (\matrix {-J & 1 & 0 & 0 & 0 \cr 0 & -J & 1 & 0 & 0 \cr
{\tilde W} & {\tilde T} & -J & {\tilde F}^{-} & G^{-} \cr 0 & 0 & 0 & -{ 3
\over 2 } J & 1 \cr {\tilde F}^{+} & G^{+} & 0 & {\tilde S} & -{ 3 \over 2 } J
\cr } \right ) \ ,}
where ${\tilde W}$, ${\tilde T}$, ${\tilde S}$ and ${\tilde F}^\pm$ denote the
currents $W$, $T$, $S$ and $F^\pm$ up to mixings with total derivatives and
polynomials of currents of lower spin. Comparing the two gauge choices we
immediately see that $G^{-}= \lambda_2^{\dagger } $ and
$F^-=\lambda^{\dagger}_1+...$. This will give rise to the classical version of
the super-$W_3$ algebra. This algebra, is precisely the \nex2 super-$W_3$
algebra that was obtained in the last section. Since we have already presented
the quantum version of this algebra we will not write down the classical
version.

The screening currents are obtained from the screening currents of the
$SL(3|2)$ current algebra via hamiltonian reduction.
Since the Borel sub-algebra is generated by three bosonic roots and one
fermionic root we have the following  screening operators:
\eqn\screenn{\eqalign {S_{\alpha_1+\alpha_2} \ = & \
e^{ 2i \alpha_- \phi_1  }  \ , \qquad  \qquad S_{\alpha_3+\alpha_4} \ =
\ e^{ 2i \alpha_-  \phi_2 } \ \cr  S_{-(\alpha_2 + \alpha_3 + \alpha_4)}
\ = & \ (\lambda^{\dagger}_2 +...) e^{ -i \alpha_- (\phi_2 + i \phi_3)}
\ \ \quad \ { \rm and}  \quad S_{\alpha_2+\alpha_3}
 \ =  \ e^{ 2i \alpha_-\phi_4 } \ .\cr }}

It is now clear how one can generalize this construction.
The constraints  are non-trivial for the currents
in the lower-right $n \times n$ corner. These non-trivial constraints coincide
with the free field realization of the nilpotent subalgebra of a $SL(1|n-1)$
sub-matrix.   The free field realization contains   $(n-1)$ pairs fermionic
fields with conformal dimension $(-n+{3 \over 2}, n-\half), \ldots, ( -\half,
{3 \over 2 } )$. As before the field with conformal dimension $-n+{3 \over 2}$
does not appear in the constraints, but will appear as a parameter in the gauge
transformation.  Following our previous analysis it is not hard to see that the
reduction gives rise to free field realization of super $W_n$-algebra  where
the  fermionic currents $F^- _i$, $(i=1,2,...(n-1))$ are realized linearly in
terms of free fermionic fields, modulo some quantum deformations.
Unfortunately, the explicit form of the currents in the chiral algebra is
complicated and not very illuminating. However, this construction of the
super-$W$ algebra serves as a proof of the existence of BRST currents for any
$W_n$-matter system coupled to the corresponding $W$-gravity.

%%%%%%%%%%%%%%%%%%%%%%%%%%%%%%%%%%%%%%%%%%%%%%%%%%%%%%%%%%%%%%%%%%%%%
%%%%%%%%%%%%%%%%%%%%%%%%%%%%%%%%%%%%%%%%%%%%%%%%%%%%%%%%%%%%%%%%%%%%%
%%%%%%%%%%%%%%%%%%%%%%%%%%%%%%%%%%%%%%%%%%%%%%%%%%%%%%%%%%%%%%%%%%%%%

\chap{Chiral rings and physical states}

It is well-known that the physical states of the matter-gravity system
are given by the non-trival cohomology of  the operator $\qbrs$.
We have seen that the improved \brs current is precisely one
of the supercharges, $G^+(z)$, of a topologically twisted
\nex2 \sc algebra. Thus one expects that the computation
of physical states to be very similar to the computation of the chiral ring,
$\cR$, in \nex2 \sc theories. However, one should note that there are several
different definitions of the chiral ring, which are only equivalent for unitary
theories \LVW:
\item{(i)} The original definition, which was to require a field $\psi\in \cR$
be both chiral ($G^+_{-1/2}\psi=0$) and primary.
\item{(ii)} The cohomology of the operator $G^+_{-1/2}$.
\item{(iii)} The fields $\psi$ that obey $h_\psi=\shalf q_\psi$, where $h_\psi$
and $q_\psi$ are the untwisted conformal weight and $U(1)$ charge.
Equivalently,
the conformal weight vanishes in the topologically twisted theory.

We will adopt $(ii)$ as our general definition of the ``chiral ring''\foot{The
ring  multiplication may not be as simple as in the unitary case.}. As will be
seen below, chiral primaries may form only a subset of $\cR$.

With this definition, all physical states of the non-critical $W$-string
system, including discrete states \multref\LZ\BMP\EXTRA,
are elements of $\cR$. We will however
concentrate on the ground ring and the tachyons, as only these fields are
present in all the ordinary gravity theories with $c^M\leq1$.

\sect{Ordinary gravity coupled to $c\leq1$ matter}

We start with the ground ring \Witgr, which consists of the physical operators
with vanishing ghost number. This ring is generated by\foot{For reasons that
will soon become clear, we use the non-standard notation $x$ and $\g$ for the
ground ring generators.}
$$
\eqalign{
x\ &=\ \big[ bc -
{t\over\sqrt{2t}}(\del\phi_L-i\del\phi_M)\big]\,V^L_{1,2}V^M_{1,2}\ ,\qquad\ \
V^L_{1,2}V^M_{1,2}\ \equiv\ e^{-\coeff 1{\sqrt{2t}}(\phi_L+i\phi_M)}\cr
\g\ &=\ \big[ bc -
{1\over\sqrt{2t}}(\del\phi_L+i\del\phi_M)\big]\,V^L_{2,1}V^M_{2,1}\ ,\qquad\ \
V^L_{2,1}V^M_{2,1}\ \equiv\ e^{-\coeff t{\sqrt{2t}}(\phi_L-i\phi_M)}\!\! .}
\EQN\gx
$$
One can represent the ring generators equivalently by replacing the matter
vertex operators by their duals so that (essentially) the holomorphic and
antiholomorphic combinations of the bosons are exchanged:
$$
\eqalign{
V^L_{1,2}V^M_{-1,-2}\ &\equiv\ e^{-\coeff 1{\sqrt{2t}}(\phi_L-i\phi_M)}
e^{i\coeff 2{\sqrt{2t}}(1-t)\phi_M}\ \cr
V^L_{2,1}V^M_{-2,-1}\ &\equiv\ e^{-\coeff t{\sqrt{2t}}(\phi_L+i\phi_M)}
e^{i\coeff 2{\sqrt{2t}}(1-t)\phi_M}\ \ .
}\EQN\dualgx
$$
We will denote the corresponding dual representatives of the ring generators
by $\hat x$ and ${\hat\gamma}^0$, respectively.

The generator $x$ is a primary field with respect to the \nex2 \sca and
has $U(1)$ charge  $q_x=\coeff1t$. The generator $\g$, on the other hand, has
charge
$q_\g=1$ and is, in general, not primary with respect to the \nex2 \sca.
Indeed we find:
$$
\g\ =\ -\big( J_{-1}+\coeff1tL_{-1}\big)\,\gamma\ ,\qquad\ \ \ \ \
\gamma\ \equiv\ \,e^{-\coeff t{\sqrt{2t}}(\phi_L-i\phi_M)}
\ , \EQN\gamdesc
$$
where $J(z)$ is given in \mguone. (It is amusing to note that for $t=1$, which
corresponds to $c^M=1$ matter coupled to gravity, $\g$ is, in fact, also
primary.)  This means that the ring of chiral primaries, which is
given by powers of $x$, is a subset of ${\cal R}$, since ${\cal R}$ also
includes powers of $\g$.

We saw in section 3 how the \brs operator that describes the coupling of matter
to gravity can be constructed by hamiltonian reduction. The models obtained by
hamiltonian reduction correspond to particular reduced \nex2 \sc models whose
free field spectrum is reduced by a Felder \brs complex. Thus, if we wish to
make a connection between the matter-gravity models and these reduced \nex2 \sc
models, we certainly must take this Felder \brs reduction into account. In the
reduced \sc models, there are two bosonic screening operators of the form
\screen\ that are easily seen to correspond precisely to the screeners
\matscreen\ of the matter model. More interesting is the additional fermionic
screener \screen, which gives rise to the extra Felder \brs operator
$$
\qt\ =\
 \oint\coeff{dz}{2\pi i}\,b(z)\,\gamma(z)\ ,
\EQN\qtilde
$$
with $\{\qt,\qbrs\}=0$.  It is elementary to see that
$$
\g(z) =\ \big\{\qt\,,\,\Gamma(z)\big\}\ ,
\EQN\gqt
$$
where $\Gamma(z)$ is the (string) \brs invariant primary field
$$
\Gamma(z)\ =\ -\coeff{(t+1)}t(\del c)(z) -
\coeff1{\sqrt{2t}}(c\,\del\phi_L)(z)\ .
\EQN\gamdef
$$
The complete Felder \brs operator of the \nex2 superconformal model is the sum
of $\tilde Q$ and a pure matter screening operator.  Since the operator
$\Gamma$ has no matter components it is annihilated by the matter screening
operator.  Therefore the observation that $\g$ is $\tilde Q$ exact implies
that it is exact with respect to the complete Felder \brs operator of the \nex2
\sc theory.  This means that the ground ring generator $\g$ is not a physical
operator of the reduced \nex2 \sc models.

Naively one might also conclude that the ground ring generator, $x$, decouples
from the \nex2 superconformal model. Indeed one also can represent $x$ as
$\tilde Q$ exact
$$
x(z) =\ \big\{\qt\,,\,F(z)\big\}\ ~~,~~~~F(z)=c\, \exx{\coeff{(t+1)}
{\sqrt{2t}}(\phi_L-i\phi_M)-\coeff2{\sqrt{2t}}\phi_L}~.
\EQN\xexact
$$
However, this does not mean that it is exact with respect to the complete
Felder \brs operator.  The matter part of the complete operator acts
non-trivially upon $F$, and so the only conclusion one obtains from \xexact\ is
that $x$ is equivalent to something else modulo Felder \brs.

We thus see that one has two basic options for defining physical models:
One can construct the reduced \nex2 \sc models, in which one employs $\qt$ as
an additional component of the Felder \brs operator, and where the only
physical ground ring generator is $x$. Alternatively, one does not employ
$\qt$, in which case one obtains matter coupled to gravity with an additional
ground ring generator $\g$.

The foregoing is parallel to the zero-momentum dilaton in the bosonic string.
The dilaton takes the form $\{\qbrs,(c_0-\bar c_0)\}\ket0$, but one does not
consider it as \brs exact since one makes the physically motivated requirement
of equivariance with respect to $(b_0-\bar b_0)$. That is, one only allows
states in the Hilbert space that are killed by $(b_0-\bar b_0)$. In particular,
$(c_0-\bar c_0)\ket0$ is disallowed, which means that the dilaton is not the
\brs variation of a physical state \doubref\DiNe\VVtopgr.

In the present context, the point is that the requirement of not using
$\qt$ as an additional \brs operator is equivalent to the requirement of
equivariance with respect to $b_0$. Indeed, in both formulations one does not
consider $\{\qt, c_0 X\}$ as an exact state. Note in particular that
$\Gamma(z)$ in \gamdef\ contains $c_0$ explicitly, which means that $\g(z)$
is not \brs exact. Note also that if one imposes the condition that
$b_0$ annihilates all physical states, the physical states must have
vanishing conformal weight, since $L_0=\{\qbrs,b_0\}$.   Thus if one makes the
additional requirement of equivariance, definition $(ii)$ of the chiral ring
implies that $(iii)$ is also satisfied.

We now briefly turn to the tachyons, which are the physical operators with
ghost number equal to one. States with non-zero ghost number fall into modules
of the ground ring \doubref\Witgr\modul. In particular, the tachyons can be
written as ground ring elements acting on certain ghost number one vacuum
states $\cP$. One must remember, however, that in free field formulations,
particular representations of physical operators act only on particular spaces
in a well-defined way. For the case at hand, this means that the ground ring
generators can be taken to act as
$$
T_{s+tr}\ \sim\ (\hat x)^s (\g)^r \cP
\EQN\tachp
$$
where ``$\sim$'' means equality up to appropriate screenings and up to \brs
exact terms. (Equivalently, the dual representatives, $x^s({\hat\gamma}^0)^r$,
can be taken to act on a dual vacuum state.) In \tachp, $\cP$ is chosen so that
all physical tachyons can be generated in this way. For minimal models
$\cM_{p,q}$, this requires taking:
$$
{\cal P} \ =\ c\,e^{\coeff 1{\sqrt{2pq}}(q-p-1)(\phi_L-i\phi_M)}\,
\exx{\sqrt{\coeff{2p}q}\phi_L}
\ ,\EQN\pdef
$$
which corresponds to the dressed matter operator with
the lowest dimension. Note also that the \lv\ momentum of $\cP$
just saturates Seiberg's bound \Sei\ of allowed \lv\ momenta,
$$
\a^L\ <\ {1\over\sqrt2}\b_0\ \equiv\ {t+1\over \sqrt{2 t}}\ ,  \EQN\seibound
$$
so that the ground state operators indeed serve as natural vacuum states in the
ghost number one sector.

For general $t$, it seems that the \nex2 \sc structure does not provide much
additional insight in the matter models coupled to gravity. However, for the
special matter models $\cM_{1,q}$, coupled to gravity, there exists a direct
connection with the \nex2 minimal models. In section 2 we noted that for
$t=q=k+2$, $k=0,1,2\dots$, the anomaly of the twisted \nex2 algebra \tneqtwo\
becomes equal to the central charge of the unitary \nex2 minimal models,
$c^{N=2}={3k\over k+2}$. This means, for integer $t\geq2$, the reduced models
are {\it identical} to the topologically twisted \nex2 minimal models, and one
can immediately infer various properties such as the structure of the chiral
ring. (For general $t=q/p$, on the other hand, the level $k$ becomes
fractional, which corresponds to non-minimal, non-unitary \nex2 superconformal
models. Not much can be said about such theories without further
investigation.) Without performing any computation, we know that the chiral
ring is characterized by the vanishing relation
$$
x^{k+1}\ =\ 0\ .\EQN\vanrel
$$
The operators $x^j$, $j=0,\dots,q-2$, are precisely the (dressed) primary
fields of the {\it minimal} $\cM_{1,q=k+2}$ models. However, the spectrum of
primary fields is not uniquely defined for $\cM_{1,q=k+2}$ models and there are
reasons for including operators outside the Kac table \doubref\kac\ind. These
operators naturally appear in two-dimensional gravity. In order to introduce
such operators, one needs to consider the equivariant cohomology. Then $\g$
becomes a physical operator, and it appears that \vanrel\ remains true \ind.
Hence the chiral ring is extended to:
$$
\cR\ =\ \{\,(x)^j(\g)^n\ |\, 0\leq j\leq k\equiv t-2,\ n\geq0\,\}\ .
\EQN\fullring
$$
This spectrum of physical operators is the same as that of topological minimal
models coupled to topological gravity \Keke, and it was indeed long suspected
that these theories should be equivalent to the $\cM_{1,q}$ models coupled to
gravity. In this context, the powers of $\g$ should be interpreted as
gravitational descendants, so that \fullring\ might be called a {\it
gravitationally extended} chiral ring of the matter model.

Indeed, for $t=2$ or $k=0$, the above model turns precisely into Distler's
formulation of pure topological gravity \dis, where $\gamma$ in \gamdesc\ plays
the role of a commuting ghost with spin $-1$. Moreover, the operators
$\g=\del\gamma+\dots$ and $\Gamma$ of \gqt, \gamdef\ correspond to the
operators denoted by $\gamma_0$ and $c_0$ in the formulation of topological
gravity of ref.\ \VVtopgr. It is known that in topological
gravity \refs{\Witopgr{,}\VVtopgr{,}\BMPtop}, $\g$ is trivial in ordinary
cohomology and hence the theory is empty. However, a non-trivial theory is
obtained when one employs equivariant cohomology.

For the $\cM_{1,q}$ models coupled to gravity, $\cP$ can thus be interpreted
as the puncture operator of topological gravity \doubref\Witopgr\VVtopgr\
$$
\cP\ \equiv\ c V^L_{0,1} V^M_{0,-1}\ =\ \ c\,\d(\gamma)\ ,
\EQN\puncdef
$$
and the extra \brs operator $\qt=\oint b\gamma$ can be interpreted as a
supersymmetry charge \doubref\dis\VVtopgr.
Indeed, the (unperturbed) three-tachyon correlation function on the sphere can
be written in the form
$$
\langle\ \cP\,\cP\,(\cP x^k)\ \rangle\ =\ 1\ ,
\EQN\correl
$$
and this is precisely the same as the correlation function that one considers
in topological gravity coupled to level $k$ topological minimal matter \Keke.

An important point is that for the $\cM_{1,q}$ models, the puncture
operator $\cP$ has vanishing \nex2 $U(1)$ charge, as well as vanishing
conformal dimension. This means that from the \nex2 point of view, the puncture
operator is on equal footing with the identity operator of the ground ring,
$\bf 1$. In other words, $\cP$ and $\bf 1$ are just different copies of the
vacuum. It is well-known that in free field realizations, physical operators
typically appear in $M$ different copies, where $M$ is the order of the
relevant Weyl group. Usually such different copies of the vacuum are given by
simple vertex operators, and this is also true in the non-standard realization
of the \nex2 superconformal algebra that we use, where one copy of the vacuum
just happens to be a tachyon with non-zero ghost number.

The discussion above implies that tachyons $(\hat x)^s(\g)^r\cP$ and ground
ring elements $(\hat x)^s(\g)^r$ have the same properties with respect to the
\nex2 superconformal algebra, and thus they should probably be interpreted as
different realizations of the same chiral ring elements. In the $\cM_{1,q}$
models coupled to gravity, the tachyons and the ground ring elements constitute
all the physical operators\foot{The ghost number two extra states are dropped
since they violate Seiberg's bound \seibound.} and therefore the ground ring
describes the full set of physical operators.

Another consequence of the realization of the \nex2 superconformal algebra in
the matter-gravity system is that it suggests that there is a much simpler way
to analyse the latter.   That is, instead of using convoluted realizations of
the \nex2 superconformal algebra like \mguone--\TG, and complicated expressions
like \gx\  to describe the ground ring, it might be better adopt the
equivalent, standard realization \bos\ \Muss\ of the \nex2 superconformal
minimal models, where, for example,  chiral ring elements can be represented
more simply by vertex operators of the form
$x^j=e^{{j\over\sqrt{2t}}(\phi_2+i\phi_1)}$.

%%%%%%%%%%%%%%%%%%%%%%%%%%%%%%%%%%%%%%%%%%%%%%%%%%%%%%%%%%%%%%%%%%%%%
%%%%%%%%%%%%%%%%%%%%%%%%%%%%%%%%%%%%%%%%%%%%%%%%%%%%%%%%%%%%%%%%%%%%%

\sect{$W_3$-gravity coupled to $c\leq2$ $W_3$-matter}

We showed in sections 2 and 3 that the (improved) \brs current of
$W_3$ strings \improv\ is part of an extended \nex2  \sc $W_3$ algebra.
This means that, as for ordinary strings, the physical operators of $W_3$
gravity coupled to matter constitute a chiral ring, $\cR$.

In ordinary strings with $c^M\!\leq\!1$ there are, typically, infinitely many
physical operators \refs{\LZ{-}\EXTRA}, at arbitrary ghost numbers. One
certainly expects to find a similar structure of extra states for
$W_3$-strings, though,
as yet, there exists no rigorous, general proof for the existence of such
states. What we will do, therefore, is to prove the existence of some of these
states by constructing them explicitly\foot{Extra states for a different class
of $W_3$ strings have recently been obtained in \Popring.}, guided by analogy
to $c^M\!\leq\!1$ strings. However, in order to avoid disrupting the line of
thought with technicalities, we will relegate the explicit construction to
Appendices A and B, and use only the results here.

We find that the analogues of tachyons and ground ring elements of ordinary
gravity are physical operators $\oz Z$, $Z=A,B,\dots,F$, with ghost numbers
ranging from to $0$ to $3$. (As in ordinary gravity, some of these states only
exist at discrete values of the momenta.) In particular, the tachyons, which
have ghost number three, are denoted by $\oz A$, and the ground ring elements
that have vanishing ghost number are denoted by $\oz F$. The ground ring is
generated by four elements, which can be represented as:
$$
\eqalign{
x_1\ &\equiv\ \oz F_{1,1;2,1} \ ,\qquad\ \ \
\gn_1\  \equiv\ \oz F_{1,2;1,1} \ ,\cr
x_2\ &\equiv\ \oz F_{1,1;1,2} \ ,\qquad\ \ \
\gn_2\  \equiv\ \oz F_{2,1;1,1} \ .\cr }
\EQN\groundgen
$$
Their explicit form is given in Appendix B. The generators have charges
$q_{x_i}=i/t$, $q_{\gn_i}=i$ with respect to the \nex2, $U(1)$ current \jone.
In
analogy to ordinary gravity, one finds that $x_1$, $x_2$ are primary with
respect to the \nex2 super-$W_3$ algebra, while $\gn_1$ and $\gn_2$ are not
primary (except when $t=1$, which corresponds to $c^M=2$ matter coupled to
$W_3$ gravity).

The associated reduced \nex2 \sc models that were obtained via hamiltonian
reduction
in section 3, are defined by employing the full set of screening operators.
This set of screening operators  \screenn\ consists of bosonic
screeners, $\oint S^\pm_M(\a_i)$, $i=1,2$ and $\oint S^\pm_L(\a_2)$,
as well as of one fermionic screener:
$$
\qt = \oint\!\coeff{dz}{2\pi i}\Big[
b_{2}-\sqrt{2} b_{1} \del\phi_{M,2}-
2i\sqrt{\coeff{t}{3}} b_{1} b_{2} c_{2} +
i\coeff{ \left( 3 + t \right) }{2{\sqrt{3t}}} (\del b_{1}) -
  \coeff{t}{3} (\del b_{1}) b_{1} c_{2}  \Big]
V^L_{12;11} V^M_{12;11} .\EQN\wqtilde
$$
All these screeners (anti-)commute with $\qbrs$.

It is straightforward to check that, up to factors and exact terms:
$$
\gn_1 \ =\ \{\qt, \Gamma_1 \}\ ,
$$
where
$$
\eqalign{
&\ \Gamma_1\ =\
  \Big( \coeff{3 i  t (9 {t^2} - 2 {t^3}+1)}{2  \left( 1
       + t \right) } \sqrt{\coeff23} \del\phi_{L,1}
     -i \coeff{ t (5 t + 4 {t^2}-5)}{\sqrt2} \del\phi_{L,2}\Big)
    \big[b_{1} (\del c_{2}) c_{2} - c_1 \big]
   \cr&+
  t \left( 1 + t + {t^2} \right) \Big(\sqrt6
      \del\phi_{M,1} + 2 {\sqrt{2}} \del\phi_{M,2} \Big)
   \big[c_{1} + b_{1} (\del c_{2}) c_{2}\big]
    \cr&+\!\!
   \Big( \coeff{t (t-9)}{4} {{\del\phi_{L,1}}^2}\! +\!
     \coeff{3 t ( 2 {t^3}-9 {t^2}-1)}{{\sqrt{3}} \left( 1 + t \right) }
      \del\phi_{L,1} \del\phi_{L,2} \!- \!
     \coeff{t (19 t + 16 {t^2}-11)}{4} {{\del\phi_{L,2}}^2}\! - \!
     \coeff{t (t-9)}{4} {{\del\phi_{M,1}}^2}
     \cr&+\!\!
     \coeff{6 t \left( 1 + t + {t^2} \right) }{{\sqrt{3}}} \del\phi_{M,1}
      \del\phi_{M,2}\!-\! \coeff{t }{4} \left( 7 + 17 t + 16 {t^2} \right)
      {{\del\phi_{M,2}}^2}\! -\!
       \coeff{ \sqrt{2 t} }{4} (t+1)(t-9)(\del^{2}\!\phi_{L,1})
 \cr&+
     \coeff{\sqrt{6t}}{4} \left( 3 + t + 4 {t^2} \right)   (3 t-1)
      (\del^{2}\!\phi_{L,2}) -
      i\coeff{\sqrt{2 t} }{4}(t-1)(t-3) (\del^{2}\!\phi_{M,1})
 \cr&-
     i\coeff{ \sqrt{6 t} }{4 } \left( 1 + 3 t \right)(t-1)
     \left( 3 + 4 t \right)(\del^{2}\!\phi_{M,2}) \Big) c_{2}+
  \Big( \coeff{3 \sqrt{2 t} }{4}( 2 {t^3}-9 {t^2}-1) \del\phi_{L,1}
\cr&-
     \coeff{3  \sqrt{6 t} }{4}\left( 1 + 3 t \right) (3 t-1) \del\phi_{L,2} -
\coeff{3 i\sqrt{2 t} }{2}(t-1)(1+t+t^2)\del\phi_{M,1} \Big)  (\del c_{2})
\cr&+
  18i(1-t)(1+t+t^2)\sqrt{\coeff t3}\big[b_{1} (\del^{2}\!c_{2}) c_{2}+(\del
b_{1})    (\del c_{2}) c_{2}\big]
\cr&+
 \coeff t2(1-9t-8t^2)\big[(\del b_{1}) c_{1} c_{2}+
 b_{1} c_{1} (\del c_{2}) + 3 b_{2} (\del c_{2}) c_{2} \big]
  }
\EQN\wgamdef
$$
is a string-$BRST$ invariant operator.  Once again this means that $\gn_1$ is
exact with respect to the complete \nex2 superconformal Felder \brs charge.
While it is not so simple to explicitly demonstrate, it also turns out that
$\gn_2$ is similarly exact.  The easiest way to see that it must be exact
follows from the connection with the topologically twisted Kazama-Suzuki models
that will be discussed below. Consequently, exactly as in ordinary gravity,
the ring generators $\gn_i$ are not physical operators in the reduced \nex2
superconformal models. However, by requiring equivariance, these operators
become physical and thus the \nex2 models describe $W_3$-matter coupled to
$W_3$-gravity.

We noted in section 2 that for the special $W_3$ minimal models
$\cM^{(3)}_{1,q=k+3}$ coupled to $W_3$ gravity, the anomaly of the topological
twisted \nex2 superconformal algebra is equal to $c^{N=2}=\coeff{6k}{k+3}$.
This is the same as the central charge of the well-known \nex2 superconformal
models of Kazama and Suzuki \KS, based on cosets $SU(3)_k/U(2)$, which happen
to have the \nex2 \sc $W_3$ algebra as their chiral algebra. This implies that
for integer $t\geq3$, the reduced models are identical to the topological
minimal $W_3$ models, whose chiral rings are given by certain polynomials in
$x_1$ and $x_2$ and have been thoroughly studied and are well understood
\doubref\LVW\cring. If one requires equivariance then the generators $\gn_1$
and $\gn_2$ also become physical, and thus the full chiral ring of the
$\cM^{(3)}_{1,q=k+3}$ model coupled to $W_3$ gravity is
$$
\cR\ =\ \{\,(x_1)^{j_1}(x_2)^{j_2}(\gn_1)^{n_1}(\gn_2)^{n_2}\ |\ 0\leq j_1+j_2\
\leq k\equiv
t-3,\,\ n_1,n_2\geq0 \,\}\ .
\EQN\wfullring
$$
One may suspect that such ``$W$-gravitationally extended'' rings of the
Kazama-Suzuki models describe the coupling of topological minimal $W_3$ models
to some form of topological $W_3$ gravity. However, the precise connection to
the kind of topological $W$-gravity that is discussed in \topw\ is
unclear, and in order to more firmly establish these statements, a deeper
analysis is required.

There are other physical operators with non-vanishing ghost numbers and these
appear to fall into modules of the ground ring. (Note that for
the $\cM_{1,q}^{(3)}$ models, the embedding diagram degenerates to a hexagon,
and hence there is only a finite number of extra ghost sectors). Adopting a
slightly different notation as in Appendix A, one thus would have, up to \brs
exact pieces:
$$
\oz Z_{s_1+tr_1,s_2+tr_2}\ \sim\
(x_1)^{s_1}(x_2)^{s_2}(\gn_1)^{r_1}(\gn_2)^{r_2}\,\cP\zl\ ,
\qquad\ \  Z=A,B,\dots,F\ .
\EQN\wmodules
$$
(If one uses a free field formulation, one must remember to perform appropriate
screenings or choose appropriate representatives of the operators $x_i$ and
$\gn_i$ in this equation.)  The operators $\cP\zl$ denotes a collection of
puncture operators, which can be represented as\foot{We have not been able to
construct the operators ${\cal P}^{(D)}$ and ${\cal P}^{(E)}$ since they do not
seem to have a form that fits a simple ansatz.}:
$$
\eqalign{
\cP^{(A)}\ &=\ \xc V^L_{0,0;1,1} V^M_{0,0;-1,-1}\cr
\cP^{(B)}\ &=\  \big[c_1c_2+\coeff i6\sqrt{\coeff3t}(3-t)(\del c_2)c_2\big]\,
V^L_{1,0;1,1-t} V^M_{1,0;1,t-2} \cr
\cP^{(C)}\ &=\  \big[c_1c_2+\coeff i6\sqrt{\coeff3t}(t-3)(\del c_2)c_2\big]\,
V^L_{1,1;1-t,1} V^M_{1,1;t-2,1} \cr
\cP^{(F)}\ &=\ {\bf 1}\ .
}\EQN\wpunc
$$
One can check that these puncture operators have vanishing quantum numbers
under the twisted \nex2 super-$W_3$ algebra, so that they should represent
equivalent copies of the vacuum. This then would imply, assuming that the
ground ring acts faithfully on these vacuum sectors, that the operators $\oz
Z$, $Z=A,B,\dots,F$, represent the same chiral ring elements.

The (unperturbed) three-point
function on the sphere can be written as
$$
\langle\ \cP^{(B)}\,\cP^{(A)}\,(\cP^{(A)} \oz F_{1,1;t-2,1})\ \rangle\ =\ 1\ ,
\EQN\correl
$$
and this is precisely what one would have for topological $W_3$ gravity
coupled to topological $W_3$ matter. In particular, all zero modes of the
ghosts on the $W_3$-sphere are saturated.
Above, $\oz F_{1,1;t-2,1}\sim(x_2)^{t-3}$
denotes the top element of the chiral ring of the topological matter model.
%Note that the correlator \correl\ is, to our knowledge, the
%first non-vanishing correlator in $W_3$-gravity that has been written down.

{}From section 3 we know that the \brs current for $W_n$-gravity coupled to
$W_n$-matter, for general $n$, is part of an \nex2 super-$W_n$ algebra. We
therefore expect that the results of this sections will generalize to arbitray
$n$. In particular, the ``topological $W_n$-strings'' with $t=k+n$,
$k=0,1,2\dots\ $, should correspond to the Kazama-Suzuki models based on
$SU(n)_k/U(n-1)$ coupled to (some form of) topological $W_n$-gravity \topw, as
conjectured in \BLNW. The ``chiral ring,'' ${\cal R}$, will be generated by
$x_i$ and $\gn_i$, $i=1,\dots,n-1$, where the $x_i$ are the usual chiral
primary fields with \nex2, $U(1)$ charges $q_{x_i} = i/t$ and the $\gn_i$ have
\nex2, $U(1)$ charges $q_{\gn_i} = i$, but are non-primary in the \nex2
superconformal algebra.

%%%%%%%%%%%%%%%%%%%%%%%%%%%%%%%%%%%%%%%%%%%%%%%%%%%%%%%%%%%%%%%%%%%%%
%%%%%%%%%%%%%%%%%%%%%%%%%%%%%%%%%%%%%%%%%%%%%%%%%%%%%%%%%%%%%%%%%%%%%

\sect{The cosmological constant and $G/G$ models}

Another approach to ``topological $W_n$-strings'' has been made by considering
$G/G$ coset models \doubref\Witgg\isr. This approach is probably equivalent to
ours, and a direct connection can be made as follows. It was observed in
\refs{\WL{,}\EY{,}\NW}\ that topological $G/G$ models (without coupling to
topological gravity) are closely related to the topologically twisted
Kazama-Suzuki models based on $G/H$. In particular, if one makes a very
specific, supersymmetry preserving perturbation of the twisted Kazama-Suzuki
model one obtains precisely the topological $G/G$ models \NW. This perturbing
operator is of the form $\psi=G^-_{-1/2}\bar G^-_{-1/2}\Phi$, where $\Phi(z)$
is a specific chiral primary field of dimension $h=\coeff k {2(k+g)}$. In our
formulation of the \nex2 \sc models, where $G/H=SU(n)/U(n-1)$, the operator
$\psi$ can be represented by
$$
\psi\ =\ S_L^-(\a_1)\ \equiv\ e^{\coeff1{\sqrt t}\a_1\shdot\phi_L}\ .
\EQN\cosm
$$
This operator describes precisely the cosmological constant perturbation,
if one interprets the Kazama-Suzuki models as theories of $W$-matter coupled to
$W$-gravity. It has a structure very similar to the other operators
$S^-_L(\a_i)$, $i=2,\dots,n-1$,
which have vanishing \nex2 superconformal quantum numbers (in contrast to
$\psi$), and which appear as screening operators in the reduced \nex2
superconformal models.

{}From the perspective of the \nex2 \sc models, $\psi$ is a relevant,
supersymmetry preserving operator, but has non-vanishing \nex2, $U(1)$ charge.
As a result, we expect the $G/G$ models to exhibit merely \nex2 supersymmetry,
and not \nex2 superconformal invariance. Moreover, because of the close
relationship of $\psi$ to the screening currents, there will in fact be an
\nex2 supersymmetric (but not superconformal) $W$-algebra at generic values of
the cosmological constant. This is precisely because the perturbation by $\psi$
of the Kazama-Suzuki models leads to quantum integrable \nex2 Toda theories
with conserved currents associated with supersymmetric $W$-algebras
\doubref\FLMW\NW. Indeed, when the model is coupled to $W$-gravity, the proof
that the the top components of the \nex2 super-$W$ multiplets \wcurr\ provide
``off-critical'' conserved $W$-charges is exactly the same as it is for the
perturbed Kazama-Suzuki models.

Putting it slightly differently, if one introduces a cosmological constant
perturbation to the $W$-string, then the perturbation breaks \nex2
superconformal invariance down to a massive \nex2 supersymmetry algebra. As a
result, the string \brs charge receives a very simple, ``off-critical''
perturbative correction, exactly as in \doubref\FMVW\LeWa. This superalgebra
will have a $W$-extension arising from the infinite number of conserved
quantities of the underlying quantum-integrable model.

%%%%%%%%%%%%%%%%%%%%%%%%%%%%%%%%%%%%%%%%%%%%%%%%%%%%%%%%%%%%%%%%%%%%%
%%%%%%%%%%%%%%%%%%%%%%%%%%%%%%%%%%%%%%%%%%%%%%%%%%%%%%%%%%%%%%%%%%%%%
%%%%%%%%%%%%%%%%%%%%%%%%%%%%%%%%%%%%%%%%%%%%%%%%%%%%%%%%%%%%%%%%%%%%%
\chap{Final comments and speculations}

{}From the results of sections 2 -- 4, it is evident that the occurence of
\nex2 superconformal symmetry is rather general in matter models coupled to
gravity and in $W$-extensions thereof. The utility of this observation will
depend greatly upon the model under consideration.  In particular, the quality
of what one learns depends upon whether the underlying \nex2 super-W model is
minimal or not.  For the $26$-dimensional bosonic string, the \nex2 structure
will probably not improve our understanding significantly, howevever, for the
minimal models $\cM_{1,q}$ coupled to $W_n$-gravity with $q=k+n$, the \nex2
structure provides valuable new insight.

The connection between topological $W$-strings and Kazama-Suzuki models raises
a number of new questions.  For example, the Kazama-Suzuki models possess \KS\
duality symmetries that imply that models based upon the coset
$$
{SU(m + n)_k \over {SU(m)_{k+n} \times SU(n)_{k+m} \times U(1)}} \ \EQN\grass
$$
yield the same \nex2  superconformal model under all permutations of $m,n$ and
$k$.  If this duality is also a symmetry for the corresponding $W$-string, it
would, amongst other things, imply the equivalence of topological
$W_{n+1}$-strings based on ${SU(n+1)_1 \over {SU(n)_{2} \times U(1)}}$ with
ordinary topological strings based on ${SU(2)_n \over { U(1)}}$.  In addition
such a duality symmetry suggests that one can generalize the reduction
procedure of section 3 to Kazama-Suzuki models, $G/H$, that are more general
than those based on $\IC\IP_n$.  In this context it is interesting to note that
for a given group $G$, the structure of the twisted \nex2 superconformal theory
is essentially determined by the choice the \nex2, $U(1)$ current\foot{This is
only strictly true for models based on hermitian, symmetric spaces.}. In our
formulation this is reflected \doubref\WL\EY\ in the various possible choices
for the fundamental weight, $\lambda_1$, of $G$ in (the appropriate
generalization of) the $U(1)$ current \jone.

Over the past year it has also become apparent that if one formulates
topological matter models in an appropriate manner, one can also obtain the
coupling to topological gravity.  For example, topological $G/G$ models
\refs{\spieg{,}\Witgg{,}\NW}\  have a finite number of topological  ``matter''
fields.  However, if the $G/G$ models are formulated using the entire gauge,
ghost and associated BRST structure \isr, it appears that one can extract the
$W$-gravity descendants.  It has also been proposed that one can obtain
topological minimal matter coupled to gravity from the \LG potentials $W(x) =
x^{k+2}$ of the topological minimal models, by careful regularization and
proper treatment of contact terms \Loss.

Our results can also be interpreted in the same spirit: We find that the free
field realization of twisted \nex2 superconformal minimal models contains the
ingredients needed to describe the coupling to topological gravity. That is,
the choice of whether or not to use the extra component, $\qt$, of \nex2 Felder
BRST operators (or, equivalently, whether or not to impose that the requirement
that cohomology be equivariant), corresponds precisely to describing the
topological minimal matter theories either with or without the coupling to
topological gravity. We have also seen that this generalizes to topological
$W$-strings. We find that each \LG variable, $x_i$, is paired with another
non-primary, chiral ring element, $\gn_i$. Each such $\gn_i$ is supposed to
correspond to a $W$-gravitational dressing of the topological matter. Based on
the proposals in \Loss\ it is tempting to suggest that these $\gn_i$'s could be
obtained directly from the \LG potentials \refs{\LVW{,}\cring{,}\LeWa}\ of the
Kazama-Suzuki models through some form of contact terms induced by integration
over ``$W$-moduli''.

The foregoing pairing of the matter fields, $x_i$, and the $W$-gravity dressing
operators, $\gn_i$, is not unexpected from the general structure of
Kazama-Suzuki models.  It is known that in the $\IC \IP_n$ models there is a
duality between the chiral ring, ${\cal R}$, and the chiral algebra, ${\cal A}$
\LNW.  Specifically, there is a one-to-one correspondence between the
generators, $x_i$, of ${\cal R}$ and the \nex2 supermultiplets, ${\cal V}_s$
\wcurr, of $W$-generators in  ${\cal A}$.  In a  theory of topological
$W$-gravity \topw, one expects a $\gamma$-ghost for each generator of the
chiral algebra, and therefore one sees that each $x_i$ will be paired with a
$\gn_i$.

Finally, we note that the presence of an \nex2 superconformal algebra
automatically implies the existence of an infinite number of higher spin
bosonic operators in the chiral algebra. That is, in addition to any
independent super-$W$ generators, there are composite operators made from
$G^+(z)$, $G^-(z)$, $J(z)$ and $T(z)$.  It was shown in \DLP\ that any \nex2
superconformal theory with $c^{N=2} \geq 3$ can be thought of as a coset
$$
{SU(1,1) \times SO(2) \over U(1)} \ , \EQN\suoneone
$$
in which the supercharges are represented by $G^\pm(z) = \psi^\pm(z) e^{\pm i
\alpha X(z)}$, where the $\psi^\pm(z)$ are parafermions of $SU(1,1)/U(1)$, and
$X(z)$ is defined by $J(z) = i \sqrt{{c\over 3}}\,\del X(z)$.  The chiral
algebra of $SU(1,1)/U(1)$ can be formed from parafermion bilinears and form an
infinite $W$-algebra \elias. For minimal models with $c^{N=2} = 3k/(k+2)$ there
is a finite $W_k$-algebra made from composites of $G^+(z)$, $G^-(z)$, $J(z)$
and $T(z)$.  It is therefore tempting to suggest that the previously known
extended $W$-symmetries of topological matter coupled to topological gravity
should be related to the foregoing composite super $W$-generators.  One should
however note that the coset formulation \suoneone\ is not exactly that of the
black-hole metric \Witbh\ since the $SU(1,1)$ structure of the black-hole
involves only Liouville and matter fields, and is orthogonal to the ghost
Hilbert space.  The $U(1)$ factors in \suoneone\ involve mixings with the
ghost
sector.  However it seems likely that the extended $W$-symmetries of the
black-hole metric should have some nice formulation in terms of the enveloping
algebra of the \nex2 superconformal algebra.

\chap{Acknowledgements}

We thank  P.\ Bouwknegt, K.\ Li, B.\ Gato-Rivera, H.\ Ooguri,  K.\ Pilch and
C.\ Vafa for discussions. W.L. also thanks John Schwarz for generous support
of his visit to Caltech, during which time this paper was completed.
We finally thank K.\ Thielemans for providing us with the Mathematica package
OPEdefs \OPE, which was essential for many of the computations.
M.B.\ is partially supported by NSF grant PHY 87/14654 and by Packard
Fellowship 89/1624, D.N.\ and N.P.W.\ are supported in part by funds provided
by the DOE under grant No. DE-FG03-84ER40168.  N.P.W. was also partially
supported by a fellowship from the Alfred P. Sloan Foundation, and is grateful
to the Theory Division at CERN for its hospitality and support during the early
stages of this work.

%%%%%%%%%%%%%%%%%%%%%%%%%%%%%%%%%%%%%%%%%%%%%%%%%%%%%%%%%%%%%%%%%%%%%
%%%%%%%%%%%%%%%%%%%%%%%%%%%%%%%%%%%%%%%%%%%%%%%%%%%%%%%%%%%%%%%%%%%%%
%%%%%%%%%%%%%%%%%%%%%%%%%%%%%%%%%%%%%%%%%%%%%%%%%%%%%%%%%%%%%%%%%%%%%

%\vfil\eject

\appendix A{Extra states for non-critical $W_3$ strings}

In section 4 we generalized some of our observations to $W_3$-gravity coupled
to $W_3$ matter. As a comprehensive study of the physical operators in
this kind of theories has not, as yet, appeared in the literature,
we will present below some remarks on the construction of such operators,
as well as some explicit results.

In ordinary strings with $c^M\!\leq\!1$ and for a given matter primary field
$\Phi_{r,s}$, the spectrum of extra states is determined by the structure of
the embedding diagram of null states over $\Phi_{r,s}$ \doubref\LZ\BMP: each
dot of this diagram denotes a null state in the matter theory, and typically
gives rise to a physical state in the matter plus gravity system. The ghost
number is given by the number of steps it takes to reach the given dot from the
top of the diagram. (The top of the diagram corresponds to the primary field
$\Phi_{r,s}$ itself, and leads to the standard tachyon in the matter plus
gravity system.) One thus obtains, generically, infinitely many ghost sectors.
In some cases (for example, for the fields in minimal matter models
$\cM_{1,q}$), the Felder resolution becomes finite \BMPa\ and the relevant part
of the diagram consists of just two points, which can be associated with the
standard tachyon and a ground ring element. These two kinds of operators are
the most generic ones in that they exist for all completely degenerate
$\Phi_{r,s}$ of the models $\mpq$ (including the $c^M=1$ model where $p=q$).

It is the analogue of this finite subset of extra states (with zero or negative
relative\foot{With ``relative ghost numbers'' we mean ghost numbers relative to
the ghost number of the tachyon. We understand that there certainly should
exist closely related states at positive relative ghost numbers, but we will
not consider these here.} ghost numbers) in which we are presently interested.
Specifically, we will associate such extra states with the ``generic'' null
states in $W_3$ minimal models, $\mpq^{(3)}$. The relevant part of the
embedding diagram looks like a hexagon whose corners are associated to the
elements of the Weyl group $W$ of $SU(3)$ \Watts\ (typically, the hexagon is
part of an infinitely extended diagram, whose dots correspond to the affine
Weyl group). For later reference, we will labels these corners by
$A,B,\dots,F$, and denote a generic such label by $Z$:

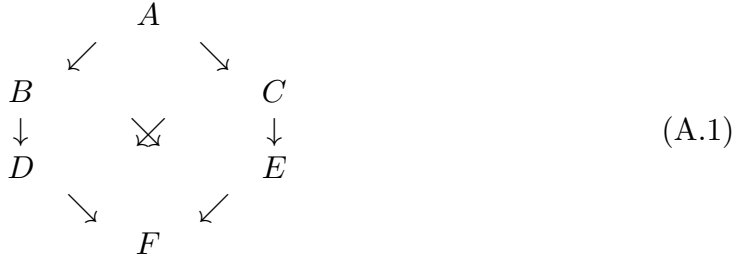
$$
\matrix{
& & A & & \cr
&\swarrow & & \searrow & \cr
B & & & & C\cr
\downarrow & & \searrow\!\!\!\!\!\swarrow & & \downarrow \cr
D & & & & E\cr
&\searrow & & \swarrow & \cr
& & F & & \cr
}\EQN\diagram
$$

The arrows correspond to embeddings of Verma modules. Label $A$
corresponds to some primary field $\prs$ (in standard notation), and labels
$B-F$ correspond to null states over this primary. The levels $\ell(Z)$ of
these states are given by:
\vskip .5cm
\begintable
sector |  element |  level    \nr
Z | $w(Z)\in W$	|      $\ell(Z)$	 \cr
A | $1	$	|	 $0$		 \nr
B | $w_1$	|      $r_1s_1$		 \nr
C | $w_2$	|      $r_2s_2$		 \nr
D | $w_2w_1$	|$\ r_1s_1+r_2s_1+r_2s_2$  \nr
E | $w_1w_2$	|$\ r_1s_1+r_1s_2+r_2s_2$  \nr
F | $w_2w_1w_2$	|$\ (r_1+r_2)(s_1+s_2)$
\endtable
\vskip .5cm
This association with Weyl group elements can be made explicit by employing a
free field formulation of the $W_3$ minimal models, as described in section 2.
Label $A$ will then be associated with the standard Fock space, $\cF_\rs$, with
$\sum r_i \leq p-1$ and $\sum s_i \leq q-1$. The Weyl group can be taken to act
on the vertex operators \vertexm\ as:
$$
w\in W:\ \ \ \a^M_{r_i;s_i}\ \to\ \a^M_{r_i;w\ast s_i} \ \equiv\ \a_+\sum_j
(r_j-1)\l_j+\a_-w(\sum(s_j-1)\l_j)\ .
\EQN\weylact
$$
The null states are then given by applying screening operators to the
transformed vertex operators, which map back into null states in the standard
Fock space. It is easy to check that under \weylact, the dimension of vertex
operators \vertexm\ changes precisely by $\ell(w(Z))$, as given in the above
table.

In analogy to usual gravity coupled to matter, we expect in $W$-gravity coupled
to matter extra \brs states to arise for each dot of the hexagon,
with ghost number $g$ given by the number of steps from the top of the the
diagram to the given dot, that is, by the length of the Weyl group element.
The top of the diagram, sector $A$, should describe the standard tachyons.
These operators have the form
$$
\oz A_\rs\ =\ \xc \,V^L_{r1,r2;-s1,-s2}V^M_\rs\ ,\EQN\tach
$$
and were shown in \BLNW\ to be indeed non-trivial elements of the \brs
cohomology. They have ghost number equal to $g=3$, and it is clear from the
hexagon that the extra states that we are seeking have ghost numbers $g=2,1,0$.
In particular, the bottom of the diagram will correspond to ground ring
elements with vanishing ghost number.

There is actually an ambiguity in writing the tachyonic (and other) operators,
as the $L_0$ and $W_0$ eigenvalues of $V^M$ and $V^L$ are invariant under
certain Weyl transformations. The Weyl group action in question, denoted by
$\Sigma$, is different from the Weyl group action $W$, defined above. For
$\Sigma$ we define:
$$
\eqalign{
\sigma_L\in\Sigma_L:\ \ \ &\a^L_{r_i;s_i}\ \to\ \a^L_{\sigma_L\ast
r_i;\sigma_L\ast s_i}\ \,\equiv \sigma_L(\a^L_{r_i;s_i}-\b_0\rho)+\b_0\rho \cr
\sigma_M\in\Sigma_M:\ \ \ &\a^M_{r_i;s_i}\ \to\ \a^M_{\sigma_M\ast
r_i;\sigma_M\ast s_i}\,\equiv \sigma_M(\a^M_{r_i;s_i}-\a_0\rho)+\a_0\rho\  \cr
}\EQN\sigmact
$$
That is, a physical operator can typically be represented by six different
vertex operators. In the matter sector, these six copies are equivalent, and we
can
always choose the standard labels, $\sum r_i \leq p-1$ and $\sum s_i \leq q-1$,
$r_i\geq1, s_i\geq1$, to represent a matter primary by a vertex operator.

The six copies in the \lv\ sector, on the other hand, are not on equal footing.
This is similar to ordinary gravity where the Weyl group is $\ZZ_2$. It is
indeed well-known \DK\ that in usual gravity there are two possible \lv\
dressings for a given matter primary $\Phi_{r,s}$; one is given by
$e^{\a^L_{r;-s}\phi_L}$ and the other one by $e^{\a^L_{-r;s}\phi_L}$. These
dressings are, however, not equivalent: the second of the dressings violates
Seiberg's condition \seibound, which requires that the \lv\ momenta must
satisfy: $\a^L < {1\over\sqrt2}\b_0$. That means that this latter dressing
should be discarded. This is extra physical input, coming from the
destabilizion of Riemann surfaces by macroscopic loops \Sei. {}From
representation theory alone, there is nothing wrong with states with the
``wrong'' dressing. They are related to the dual resolution in the Felder
complex, and give rise to extra states with zero and positive relative ghost
number. Moreover, the matching of scaling dimensions of operators in matter
plus gravity theories with those of matrix models also requires restriction to
operators that satisfy Seiberg's condition \seibound\ \Sei. Furthermore, we
introduced in section 2 certain \nex2, $U(1)$ currents that involve the \lv\
field explicitly. The $U(1)$ charge of \lv\ vertex operators will thus in
general depend on $\Sigma_L$ transformations, and this means that the \lv\
dressings cannot be equivalent.

Unfortunately, we do not rigorously know what the analogue of Seiberg's bound
is for $W$-gravity. But we note that $\a_L=\b_0\rho$ is the fixed point of the
Weyl transformations \sigmact\ and it is this what should correspond to
Seiberg's bound \seibound. We will therefore assume that the \lv\ momentum must
lie within (and not on) the boundary of a particular Weyl chamber shifted by
$\b_0\rho$. This Weyl chamber corresponds precisely to the choice that we made
in the representation \tach\ of the tachyons. (Our assumption will be justified
{\it a posteriori}, in that we will indeed find non-trivial cohomology at the
correct (negative relative) ghost numbers. A further consistency check is as
follows. For the particular weight basis:
$\l_1=(\coeff1{\sqrt2},\coeff1{\sqrt6}),\l_2=(0,\coeff 2{\sqrt6})$, it can be
inferred from \LPSX\ that $\phi_{L,1}$ can be viewed as the ``usual'' \lv\
field as part of $W$-gravity. Our choice of tachyon representative indeed
satisfies Seiberg's condition for the first components of \lv\ momenta:
$(\a^L_{r_i,-s_i})_1<\coeff1{\sqrt2}\b_0$ for labels in the standard range.)

As in usual theories of gravity coupled to matter, we expect that the \lv\
dressing of an extra state corresponds to the dressing of a null state over a
matter primary. Thus, the extra state should be given by a polynomial, $X$, in
the ghosts and in $\del\phi_{L,M}$, of appropriate conformal dimension
$\ell(Z)$ and ghost number $g$, times a vertex operator piece $\cV$:
$$
\oz Z_\rs\ =\ X\zl_{-\ell(Z)}\,\xc\,\cV\zl_\rs\ .\EQN\extra
$$
The \lv\ dressing can be obtained by Weyl transformations \weylact\ $w_L\in
W_L$ acting on the \lv\ part of the tachyons \tach:
$$
\cV\zl_\rs \equiv V^L_{r_1,r_2;w(Z)*(-s_1),w(Z)*(-s_2)}V^M_\rs\ .
\EQN\VZ
$$
{}From the first table we have
$$
h(\cV\zl) = 4-\ell(Z)\ , \EQN\hvz
$$
so that all operators $\oz Z_\rs$ have vanishing dimensions. More explicitly,
the precise \lv\ dressings are:

\vskip .5cm
\begintable
sector 	| Liouville dressing		 	|   ghost number of\nr
Z 	|$V^L_{r_1,r_2;w(Z)*(-s_1),w(Z)*(-s_2)}$	|  $X\zl$ \cr
A 	|$V^L_{r_1,r_2;-s_1,-s_2}  	$	|       0	 \nr
B 	|$V^L_{r_1,r_2;s_1,-s_1-s_2}   	$ 	|       $-1$	 \nr
C 	|$V^L_{r_1,r_2;-s_1-s_2,s_2}	$	|       $-1$	 \nr
D 	|$V^L_{r_1,r_2;-s_2,s_1+s_2}	$	|       $-2$	 \nr
E 	|$V^L_{r_1,r_2;s_1+s_2,-s_1}	$	|       $-2$	 \nr
F 	|$V^L_{r_1,r_2;s_2,s_1}		$	|       $-3$
\endtable
\vskip .5cm

All that remains is to make an ansatz for $X\zl$ with the correct
quantum numbers and to solve for $[\qbrs,\oz Z_\rs\}=0$, where the \brs current
is given by \jb \foot{Our representatives are such that $\qbrs$ maps into
vanishing null states. This faciliates the computation.}. This yields a highly
overdetemined set of equations, and the fact that we do find solutions
shows that our assumptions about the structure
of the extra states were correct. (One cannot exclude at this point the
possibility that some solutions give $BRST$-exact operators. We have checked,
however, that our solutions indeed are non-trivial with respect to \jb.)

The simplest non-tachyonic operators that we find are those whose
dimension of the vertex operator part does not depend on some
combination of the $r_i,s_i$. For example, the dimension of $\cV^{(B)}$ is
independent of $s_2\equiv s$, and we find explicitly:
$$
\oz B_{1,1;1,s}\ =\ \Big[c_1c_2+ \coeff i6\sqrt{\coeff3t}(3t-2s-1)(\del
c_2)c_2\Big]\,V^L_{1,1;1,-1-s}V^M_{1,1;1,s}\EQN\ones
$$
Similarly,
$$
\oz C_{1,1;s,1}\ =\ \Big[c_1c_2- \coeff i6\sqrt{\coeff3t}(3t-2s-1)(\del
c_2)c_2\Big]\,V^L_{1,1;-1-s,1}V^M_{1,1;s,1}\EQN\sone
$$
Slightly more non-trivial is:
$$
\eqalign{\oz C_{1,1;s,2}\ &=\ \Big[ -{i\over \sqrt{2t}} c_1 (\del c_2)
- {1\over\sqrt3}c_1 c_2[\partial\phi_{M,2}+i\partial\phi_{L,2}]\cr
&+{1\over {6 {\sqrt{t}}}} (\del c_2)c_2[ \left( 5  + 2 s - 3 t \right)
\partial\phi_{L,2} -i \left(2 s - 3 t-1\right)  \partial\phi_{M,2}]\cr
 &+ {1\over\sqrt6\, t} ( 2 t-s-1 )  b_1 c_1(\del c_2) c_2
+{1\over \sqrt6}(\del^2\!c_2) c_2 \Big]\, V^L_{1,1;-2-s,2} V^M_{1,1;s,2}
}\EQN\stwo
$$
Moreover, for sectors $D$ and $E$ we find that the simplest $BRST$ invariant
operators involve ``screening'' operators:
$$
\eqalign{
\oz E_{1,1;1,1} &=
\Big[c_1-b_1(\del c_2)c_2 + \coeff i2\sqrt{\coeff3t}(t+1)(\del c_2)
+ \coeff i{\sqrt{2}}c_2[\del\phi_{L,2}+\sqrt3\del\phi_{L,1}]\Big]\,
S_L^-(\a_2)\cr
\oz E_{1,1;-1,2} &=
\Big[c_1+b_1(\del c_2)c_2 - \coeff i2\sqrt{\coeff3t}(t-1)(\del c_2)
+ \sqrt{2}c_2\del\phi_{M,2}\Big]\,
S_M^-(\a_1)  }\EQN\Oscr
$$
where $S_{M}^\pm(\a_i)$ are the matter screening operators \matscreen\ and
$S_{L}^\pm(\a_i)$ are their \lv\ counterparts. Similarly, $\oz D_{1,1;1,1} \sim
S_L^-(\a_1)$, {\it etc}. Note that the foregoing are truly \brs invariant
operators. From the descent equations \descent\ we infer that the screening
operators themselves are \brs invariant up to total derivatives, so that $\oint
S^\pm_{L,M}(\a_i)$ are \brs invariant. Note also, as explained in more detail
in section 4.3, that one of these operators (namely $\oint S^-_{L}(\a_1)$) is
not a true screening operator, but represents a physical perturbation of the
action.

More interesting, but also more complicated, are the ground ring operators. The
dimension of $X^{(F)}$ grows rapidly with $r_i$ and $s_i$, and therefore its
complexity dramatically increases at the same time. As a result we are only
able to give the generators of the ground ring. We have collected the explicit
expressions in Appendix B. Note that these operators exist for generic values
of $t$, and in particular for $t=1$, which describes $c^M=2$ matter coupled to
$W_3$ gravity. We expect that there should be a way of writing these ground
ring generators, as well as general ground ring elements, more compactly in
terms of Schur-like polynomials, but we have not succeeded in doing this.

Note also that these operators are defined only up to $BRST$ trivial pieces of
the form $\{\qbrs,b_1 V^L V^M\}$. Using this freedom, one can remove the pieces
proportional to $\del^2\!c_2$ in the expressions for the ground ring
generators, so that all generators are killed by the zero modes of the $b$
ghosts. However, we preferred to choose different representatives, namely
operators that are primary with respect to the \nex2 \sc algebra (the
expressions for $\gn_1,\gn_2$ are primary only for $t=1$).

%%%%%%%%%%%%%%%%%%%%%%%%%%%%%%%%%%%%%%%%%%%%%%%%%%%%%%%%%%%%%%%%%%%%%
%%%%%%%%%%%%%%%%%%%%%%%%%%%%%%%%%%%%%%%%%%%%%%%%%%%%%%%%%%%%%%%%%%%%%
%%%%%%%%%%%%%%%%%%%%%%%%%%%%%%%%%%%%%%%%%%%%%%%%%%%%%%%%%%%%%%%%%%%%%

\vfil\eject

\def\append#1#2{\global\meqno=1\global\subsecno=0\xdef\secsym{\hbox{#1.}}
\bigbreak\bigskip\noindent{\bf Appendix #1. #2}\message{(#1. #2)}
\writetoca{Appendix {#1.} {#2}}\nobreak}

\append {B}{Ground ring generators for $W_3$ gravity}
$$
\eqalign{
&x_1\ \equiv\ \oz F_{1,1;2,1}\ =\ \cr
&\Big[
\coeff{2}{3} {{\del\phi_{L,2}}^2}
  - \coeff{i}{{\sqrt{3}}} \del\phi_{L,2} \del\phi_{M,1}
  - \coeff{i}{3} \del\phi_{L,2} \del\phi_{M,2} -
  \coeff{1}{{\sqrt{3}}} \del\phi_{M,1} \del\phi_{M,2} +
  \coeff{1}{3} {{\del\phi_{M,2}}^2}
  \cr&-
  \coeff{2}{3 t} b_{1} b_{2} c_{1} c_{2} -
  \coeff{5 i}{3t {\sqrt{3t}}} b_{1} b_{2} (\del c_{2}) c_{2} -
  \sqrt{\coeff2{3t}}\del\phi_{L,2} b_{1} c_{1}
  + i \coeff{(21 {t^2}-5)}{12 t{\sqrt{3t}} }(\del b_{1}) (\del c_{2})
  \cr&+
  \left( i \coeff{(t-5)}{3 {\sqrt{2}} t} \del\phi_{L,2} -
     \coeff{2}{\sqrt{6}} \del\phi_{M,1} - \coeff{\sqrt2}{3} \del\phi_{M,2}
      \right)  b_{1} (\del c_{2}) +
  i \coeff{(4 t-3)}{2 \sqrt{3 t}} b_{1} (\del^{2}\!c_{2}) +
  i \coeff{1}{\sqrt{3 t}} b_{2} c_{1}
  \cr&+
  \Big(\coeff i{ {\sqrt{2t}}}\del\phi_{M,1} -\sqrt{\coeff2{3 t}}\del\phi_{L,2}+
     \coeff i{ \sqrt{6 t}} \del\phi_{M,2} \Big)  b_{2} c_{2} +
  \coeff{(9 t-5)}{6 t} b_{2} (\del c_{2}) +
  \coeff{i}{3t {\sqrt{3t}}} (\del b_{1}) b_{1} c_{1} c_{2}
  \cr&+
  \coeff{(9 {t^2}-24t-5)}{18 {t^2}} (\del b_{1}) b_{1} (\del c_{2}) c_{2} +
  \coeff{(3 t-1)}{6 t} (\del b_{1}) c_{1}
  + \coeff i2\sqrt{\coeff t3} (\del^{2}\!b_{1}) c_{2} -
  \sqrt{\coeff{2t}3} (\del^{2}\!\phi_{L,2})
    \cr&+
  \Big( i \coeff{{\sqrt{2}} (t-1)}{6 t} \del\phi_{L,2} -
     \coeff{ \left( 1 + t \right) }{4  t}\sqrt{\coeff23} \del\phi_{M,1} -
     \coeff{{\sqrt{2}} \left( 1 + t \right) }{12 t} \del\phi_{M,2} \Big)
   (\del b_{1}) c_{2}  + (\del b_{2}) c_{2}
   \cr&+
    b_{1} c_{2} \Big( \coeff{2 i}{3 \sqrt{3 t}} {{\del\phi_{M,2}}^2} +
     \coeff{2 i}{3 {\sqrt{2}}} (\del^{2}\!\phi_{L,2}) -
     \coeff{4 (t-1)}{3 {\sqrt{2}} t} (\del^{2}\!\phi_{M,2}) -
i\coeff{2}{3 \sqrt{3 t}} {{\del\phi_{L,2}}^2}
\cr&-
     \coeff{1}{3 {\sqrt{t}}} \del\phi_{L,2} \del\phi_{M,1} -
     \coeff{1}{3 \sqrt{3 t}} \del\phi_{L,2} \del\phi_{M,2} -
     \coeff{2 i}{3 {\sqrt{t}}} \del\phi_{M,1} \del\phi_{M,2} \Big)
     +\coeff{2 i (t-1)}{\sqrt{6 t}} (\del^{2}\!\phi_{M,2})
\Big]\cr &\ \ \times\ V^L_{1,1;1,2}V^M_{1,1;2,1}
}\EQN\ringa
$$\nobreak\penalty10000$$
\eqalign{
&x_2\ \equiv\ \oz F_{1,1;1,2}\ =\ \cr
&\Big[
  \coeff{1}{3} {{\del\phi_{L,2}}^2} +  \coeff{i}{{\sqrt{3}}} \del\phi_{L,1}
\del\phi_{M,2} +
   \coeff{i}{3} \del\phi_{L,2} \del\phi_{M,2} + \coeff{2}{3}
{{\del\phi_{M,2}}^2}
  - \coeff{1}{{\sqrt{3}}} \del\phi_{L,1} \del\phi_{L,2}
 \cr&-
  i \sqrt{\coeff{2}{3t}} \del\phi_{M,2} b_{1} c_{1} +
  \coeff{2}{3 t} b_{1} b_{2} c_{1} c_{2} -
  i\coeff{5 }{3 t{\sqrt{3t}} } b_{1} b_{2} (\del c_{2}) c_{2}
   -\left( 1 + t \right)\sqrt{\coeff 2{3 t}} (\del^2\!\phi_{L,2})
  \cr&+
  \left( \coeff{2 i}{\sqrt{6}} \del\phi_{L,1} + \coeff{2 i}{3 {\sqrt{2}}}
\del\phi_{L,2} -
     \coeff{5 + t}{3 {\sqrt{2}} t} \del\phi_{M,2} \right)  b_{1} (\del c_{2}) +
  i \coeff{3 + 4 t}{2 \sqrt{3 t}} b_{1} (\del^{2}\!c_{2}) + i \coeff{1}{\sqrt{3
t}} b_{2} c_{1}
  \cr&+
  \Big( \coeff1{{\sqrt{2t}}} \del\phi_{L,1} +
     \coeff1{\sqrt{6t}} \del\phi_{L,2} -
     i \sqrt{\coeff{{{2}}}{{3 t}}} \del\phi_{M,2} \Big)  b_{2} c_{2} +
  \coeff{5 + 9 t}{6 t} b_{2} (\del c_{2}) + \coeff{1 + 3 t}{6 t} (\del b_{1})
c_{1} \cr&+
  \Big( i \coeff{ (t-1)}{4  t} \sqrt{\coeff23}\del\phi_{L,1} +
     i \coeff{{\sqrt{2}} (t-1)}{12 t} \del\phi_{L,2} -
     \coeff{{\sqrt{2}} \left( 1 + t \right) }{6 t} \del\phi_{M,2} \Big)  (\del
b_{1}) c_{2} +
  i \coeff{1}{3t {\sqrt{3t}}} (\del b_{1}) b_{1} c_{1} c_{2}
  \cr&-
  \coeff{(24 t+
  9 {t^2}-5)}{18 {t^2}} (\del b_{1}) b_{1} (\del c_{2}) c_{2} +
  i \coeff{(21 {t^2}-5)}{12t {\sqrt{3t}}} (\del b_{1}) (\del c_{2}) +
  (\del b_{2}) c_{2} + i \coeff12\sqrt{\coeff t3} (\del^{2}\!b_{1}) c_{2}
  \cr&+
  b_{1} c_{2} \Big( \coeff{2 i}{3 {\sqrt{t}}} \del\phi_{L,1} \del\phi_{L,2} -
     \coeff{2 i}{3 \sqrt{3 t}} {{\del\phi_{L,2}}^2} -
     \coeff{1}{3 {\sqrt{t}}} \del\phi_{L,1} \del\phi_{M,2} -
     \coeff{1}{3 \sqrt{3 t}} \del\phi_{L,2} \del\phi_{M,2}
     \cr&+
     \coeff{2 i}{3 \sqrt{3 t}} {{\del\phi_{M,2}}^2} +
     \coeff{4 i \left( 1 + t \right) }{3 {\sqrt{2}} t} (\del^2\!\phi_{L,2}) -
     \coeff{2}{3 {\sqrt{2}}} (\del^{2}\!\phi_{M,2}) \Big)
 + 2 i \sqrt{\coeff t6} (\del^{2}\!\phi_{M,2})
\Big]\,V^L_{1,1;2,1}V^M_{1,1;1,2}
}\EQN\ringb
$$
\goodbreak

$$
\eqalign{
&\gn_1\ \equiv\ \oz F_{1,2;1,1}\ =\ \cr
&\Big[
  \coeff{3 + 5 t}{12} {{\del\phi_{L,2}}^2}
  - \coeff{(t-1)}{4} {{\del\phi_{L,1}}^2}   +
  \coeff{1 + t}{4} {{\del\phi_{M,1}}^2} +
  \coeff{2 i t}{3} \del\phi_{L,2} \del\phi_{M,2} -
  \coeff{(5 t-3)}{12} {{\del\phi_{M,2}}^2}
  \cr&-
  \coeff{2 {t^2}}{3} b_{1} b_{2} c_{1} c_{2}-
  \coeff{5 i {t^3}}{3 {\sqrt{3t}}}
   b_{1} b_{2} (\del c_{2}) c_{2} -
  \Big( (t+1) \sqrt{\coeff t6}\del\phi_{L,2}  +
     i (t-1) \sqrt{\coeff t6}\del\phi_{M,2} \Big)  b_{1} c_{1}
   \cr&-
   i \coeff{3 + 5 {t^2}}{6 {\sqrt{2}}}
  \Big( \del\phi_{L,2} + i \del\phi_{M,2} \Big)
   b_{1} (\del c_{2}) +  \coeff i2 \sqrt{\coeff t3}
   b_{1} (\del^{2}\!c_{2}) + i \coeff{{t^2}}{{\sqrt{3t}}}
   b_{2} c_{1}   -   \coeff{({t^2}-3)}{6} (\del b_{1}) c_{1}
   \cr&-
   t^2 \sqrt{\coeff2{3t}} \Big( \del\phi_{L,2}+ i  \del\phi_{M,2}\Big)
       b_{2} c_{2} - \coeff{(5 {t^2}-9)}{6} b_{2} (\del c_{2}) +
  i \coeff{{t^{{5\over 2}}}}{3 {\sqrt{3}}} (\del b_{1}) b_{1} c_{1} c_{2}
   + (\del b_{2}) c_{2}
  \cr&-
  \coeff{5 t \left( 3 + {t^2} \right) }{18}(\del b_{1}) b_{1} (\del c_{2})
c_{2}
  -  i  \Big(  \coeff{{\sqrt{2}} t (t-1)}{6} \del\phi_{L,2} +
     i\coeff{{\sqrt{2}} t \left( 1 + t \right) }{6} \del\phi_{M,2} \Big)
   (\del b_{1}) c_{2}
   \cr&-
   \coeff{it (5 {t^2}-21)}{12 {\sqrt{3t}}}
   (\del b_{1}) (\del c_{2})  +
 \coeff{it}{2 {\sqrt{3t}}} (\del^{2}\!b_{1}) c_{2}\! +\!
  \coeff{ \left( 1 + t \right)  (t-1)}{2 {\sqrt{2t}}}
   (\del^{2}\!\phi_{L,1})\! +\! \coeff{ \left( 1 + t \right)  (t-3)}
   {2\sqrt{6 t}} (\del^{2}\!\phi_{L,2})
   \cr&+
  \Big( \coeff{it(1-t)}{2 {\sqrt{3t}}} {{\del\phi_{L,1}}^2}
     - \coeff{it \left( 3 + t \right) }{6 {\sqrt{3t}}}
      {{\del\phi_{L,2}}^2} +
      \coeff{it \left( 1 + t \right) }{2 {\sqrt{3t}}}
      {{\del\phi_{M,1}}^2} + \coeff{2 {t^2}}{3 {\sqrt{3t}}}
      \del\phi_{L,2} \del\phi_{M,2} +
     \coeff{it (t-3)}{6 {\sqrt{3t}}} {{\del\phi_{M,2}}^2}
     \cr&+
     i \coeff{ \left( 1 + t \right)  (t-1)}{{\sqrt{6}}}
      (\del^{2}\!\phi_{L,1}) +
     i \coeff{t \left( 1 + t \right) }{3 {\sqrt{2}}}
      (\del^{2}\!\phi_{L,2}) -
     \coeff{\left( 1 + t \right)  (t-1)}{\sqrt{6}} (\del^{2}\!\phi_{M,1}) -
     \coeff{t (t-1)}{3 {\sqrt{2}}} (\del^{2}\!\phi_{M,2}) \Big) b_{1} c_{2}
   \cr&+
  i \coeff{ \left( 1 + t \right)  (t-1)}{2 {\sqrt{2t}}}
   (\del^{2}\!\phi_{M,1}) + i \coeff{\left( 3 + t \right)  (t-1)}{2
   \sqrt{6t}} (\del^{2}\!\phi_{M,2})
\Big]\,V^L_{1,2;1,1}V^M_{1,2;1,1}
}\EQN\ringc
$$\nobreak
$$
\eqalign{
&\gn_2\ \equiv\ \oz F_{2,1;1,1}\ =\ \cr
&\Big[\coeff{3 + t}{12} {{\del\phi_{L,2}}^2}
-\coeff{(t-1)}{4} {{\del\phi_{L,1}}^2}  -
  \coeff{t}{{\sqrt{3}}} \del\phi_{L,1} \del\phi_{L,2}
  -i \coeff{t}{2} \del\phi_{L,1} \del\phi_{M,1}  +
  \coeff{t}{{\sqrt{3}}} \del\phi_{M,1} \del\phi_{M,2}
  \cr&-
  i \coeff{t}{2 {\sqrt{3}}} \del\phi_{L,2} \del\phi_{M,1} +
  \coeff{1 + t}{4} {{\del\phi_{M,1}}^2}
  -i \coeff{t}{2 {\sqrt{3}}} \del\phi_{L,1} \del\phi_{M,2}
  -i \coeff{t}{6} \del\phi_{L,2} \del\phi_{M,2}   \cr&-
\coeff{(t-3)}{12}{{\del\phi_{M,2}}^2}+\coeff{2 {t^2}}{3} b_{1} b_{2} c_{1}
c_{2}-
  \coeff{5 i {t^3}}{3 {\sqrt{3t}}} b_{1} b_{2} (\del c_{2}) c_{2}
     +\coeff{7 i t}{2 {\sqrt{3t}}} b_{1} (\del^{2}\!c_{2})
  \cr&+
  \Big( \coeff{(t-1) \sqrt{2 t} }{4}
      \del\phi_{L,1} + \coeff{ (t-1)
       }{4 } \sqrt{\coeff{2t}3}\del\phi_{L,2} +
      \coeff{\left( 1 + t \right) i \sqrt{2 t}}{4} \del\phi_{M,1} +
     \coeff{\left( 1 + t \right) i}{4}\sqrt{\coeff{2t}3} \del\phi_{M,2} \Big)
   b_{1} c_{1}
   \cr&+\!
   \Big( \coeff{( 5 {t^2}+10t-3)}{4 \sqrt{6}} \del\phi_{M,1}\! + \!
     \coeff{( 5 {t^2}+10t-3)}{12 {\sqrt{2}}} \del\phi_{M,2}
    \!-\! \coeff{i (5 {t^2}-10t-3)}{4 \sqrt{6}} \del\phi_{L,1}
\!-\!\coeff{i( 5 {t^2}-10t-3)}{12 {\sqrt{2}}} \del\phi_{L,2} \Big)b_{1}\del
c_{2}
    \cr&+
  \Big( t\sqrt{\coeff t2} \del\phi_{L,1} +
     \coeff{ {t^2}}{2 } \sqrt{\coeff2{3t}}\del\phi_{L,2} +
     i t\sqrt{\coeff t2} \del\phi_{M,1} +
     i \coeff{ {t^2}}{2} \sqrt{\coeff2{3t}} \del\phi_{M,2} \Big)
   b_{2} c_{2} + \coeff{9 + 5 {t^2}}{6} b_{2} (\del c_{2})
   \cr&+
  i \coeff{{t^3}}{3 {\sqrt{3t}}} (\del b_{1}) b_{1} c_{1} c_{2} +
  \coeff{t (5 {t^2}-33)}{18} (\del b_{1}) b_{1} (\del c_{2}) c_{2} +
  \coeff{(3 + {t^2})}{6} (\del b_{1}) c_{1}
    +i \coeff{{t^2}}{{\sqrt{3t}}} b_{2} c_{1}
  \cr&-
  \Big( i \coeff{ t (t-1)}{4 } \sqrt{\coeff23}\del\phi_{L,1}
     -i \coeff{{\sqrt{2}} t (t-1)}{12} \del\phi_{L,2} -
     \coeff{ t \left( 1 + t \right) }{4 }\sqrt{\coeff23} \del\phi_{M,1} -
     \coeff{{\sqrt{2}} t \left( 1 + t \right) }{12} \del\phi_{M,2} \Big)
   (\del b_{1}) c_{2} -
   }
$$
\goodbreak
$$
\eqalign{
   &-
   i \coeff{t (5 {t^2}-21)}{12 {\sqrt{3t}}}
   (\del b_{1}) (\del c_{2}) +
  i \coeff{t}{2 {\sqrt{3t}}} (\del^{2}\!b_{1}) c_{2} +
  \coeff{ (t-1)}{2 {\sqrt{2t}}} (\del^{2}\!\phi_{L,1})+ (\del b_{2}) c_{2}
  \cr&-
  \coeff{ \left( 3 + 3 t + 2 {t^2} \right) }{2 \sqrt{6 t}}
   (\del^{2}\!\phi_{L,2})  -i \coeff{ \left( 1 + t \right) }{2 {\sqrt{2t}}}
   (\del^{2}\!\phi_{M,1})  -
   i \coeff{3 - 3 t + 2 {t^2}}{2 \sqrt{6 t}}
   (\del^{2}\!\phi_{M,2})+
   b_{1} c_{2}
   \Big(     \coeff{{t^{{3\over 2}}}}{6} \del\phi_{L,2} \del\phi_{M,1}
 \cr&-
  i \coeff{ t(t-1) }{4{\sqrt{3t}}} {{\del\phi_{L,1}}^2} +
     i \coeff{{\sqrt{t}} \left( 3 + t \right) }{6} \del\phi_{L,1}
\del\phi_{L,2}
     -i \coeff{t \left( 3 + 5 t \right) }{12 {\sqrt{3t}}} {{\del\phi_{L,2}}^2}
     \! -\!\coeff{i{\sqrt{t}} (t-3)}{6} \del\phi_{M,1} \del\phi_{M,2}
      \cr&+\!
     \coeff{{t^2}}{2 {\sqrt{3t}}} \del\phi_{L,1} \del\phi_{M,1}\! +\!
  \coeff{it \left( 1 + t \right) }{4 {\sqrt{3t}}} {{\del\phi_{M,1}}^2}\! +\!
     \coeff{{t^{{3\over 2}}}}{6} \del\phi_{L,1} \del\phi_{M,2}\! +\!
     \coeff{{t^2}}{6 {\sqrt{3t}}} \del\phi_{L,2} \del\phi_{M,2}
         \cr&+
     i \coeff{t (5 t-3)}{12 {\sqrt{3t}}} {{\del\phi_{M,2}}^2} +
     i \coeff{ (t-1)}{2 {\sqrt{6}}} (\del^{2}\!\phi_{L,1}) +
     i \coeff{3 + 9 t + 4 {t^2}}{6 {\sqrt{2}}} (\del^{2}\!\phi_{L,2}) +
     \coeff{1 + t}{2 \sqrt{6}} (\del^{2}\!\phi_{M,1})
     \cr&-
     \coeff{3 - 9 t + 4 {t^2}}{6 {\sqrt{2}}} (\del^{2}\!\phi_{M,2}) \Big)
\Big]\,V^L_{2,1;1,1}V^M_{2,1;1,1}
}\EQN\ringd
$$

%%%%%%%%%%%%%%%%%%%%%%%%%%%%%%%%%%%%%%%%%%%%%%%%%%%%%%%%%%%%%%

\refout
\end